\documentclass[acmsmall,screen]{acmart}

\usepackage{booktabs}
\usepackage[utf8]{inputenc}
\usepackage[T1]{fontenc}
\usepackage{lineno}
\usepackage{hyphenat}
\modulolinenumbers[5]
\usepackage{epstopdf}

\usepackage[scaled=.8]{beramono}
\usepackage{dsfont}
\usepackage{lineno}
\usepackage[T1]{fontenc}
\usepackage[flushleft]{threeparttable}
\usepackage{xspace}
\usepackage{amsthm}
\usepackage{xcolor}
\usepackage{lscape}
\usepackage{url}
\usepackage{multirow}
\usepackage{array}
\usepackage{paralist}
\usepackage{makecell}
\usepackage{adjustbox}
\usepackage{support-caption}
\usepackage[framemethod=tikz]{mdframed}
\usepackage{float}
\usepackage[inline]{enumitem}
\usepackage{rotating}
\usepackage{fp}
\usepackage[autolanguage]{numprint}
\newcommand*\np[2][z]{
		\ifx z#1%
		$\numprint{#2}$%
		\else%
		$\numprint[#1]{#2}$%
		\fi\xspace
}
\usepackage{parcolumns}

\usepackage{scalerel}
\usepackage{tikz}

\definecolor{orcidlogocol}{HTML}{A6CE39}
\tikzset{
	orcidlogo/.pic={
		\fill[orcidlogocol] svg{M256,128c0,70.7-57.3,128-128,128C57.3,256,0,198.7,0,128C0,57.3,57.3,0,128,0C198.7,0,256,57.3,256,128z};
		\fill[white] svg{M86.3,186.2H70.9V79.1h15.4v48.4V186.2z}
		svg{M108.9,79.1h41.6c39.6,0,57,28.3,57,53.6c0,27.5-21.5,53.6-56.8,53.6h-41.8V79.1z M124.3,172.4h24.5c34.9,0,42.9-26.5,42.9-39.7c0-21.5-13.7-39.7-43.7-39.7h-23.7V172.4z}
		svg{M88.7,56.8c0,5.5-4.5,10.1-10.1,10.1c-5.6,0-10.1-4.6-10.1-10.1c0-5.6,4.5-10.1,10.1-10.1C84.2,46.7,88.7,51.3,88.7,56.8z};
	}
}

\newcommand\orcidicon[1]{\href{https://orcid.org/#1}{\mbox{\scalerel*{
				\begin{tikzpicture}[yscale=-1,transform shape]
					\pic{orcidlogo};
				\end{tikzpicture}
			}{|}}}}

\mdfdefinestyle{mpdframe}{
	frametitlebackgroundcolor   =black!15,
	frametitlerule              =true,
	roundcorner                 =5pt,
	middlelinewidth             =0.8pt,
	innermargin                 =0.2cm,
	outermargin                 =0.2cm,
	innerleftmargin             =0.2cm,
	innerrightmargin            =0.2cm,
	innertopmargin              =0.2cm,
	innerbottommargin           =0.2cm
}

\usepackage{tabularx}
\newcolumntype{C}{>{\centering\arraybackslash}X}
\newcolumntype{R}{>{\raggedleft\arraybackslash}X}

\usepackage{booktabs}

\usepackage{xcolor}
\usepackage{hyperref}

\usepackage{float}
\floatstyle{plaintop}
\restylefloat{table}

\makeatletter
\newif\if@restonecol
\makeatother

\usepackage[linesnumbered,ruled,vlined]{algorithm2e}

\SetCommentSty{mycommfont}

\SetKwInput{KwData}{Input}
\SetKwInput{KwResult}{Output}

\newtheorem{definition}{Definition}

\usepackage{array}
\usepackage{stfloats}
\usepackage{url}
\usepackage{tabularx}
\usepackage{xspace}
\usepackage{graphicx}
\usepackage{subcaption}

\usepackage{tcolorbox,tikz}
\usepackage{soul}
\usetikzlibrary{arrows.meta}
\usetikzlibrary{arrows}
\usetikzlibrary{shapes}
\usetikzlibrary{patterns} 
\usetikzlibrary{positioning}
\usetikzlibrary{calc}
\usetikzlibrary{fpu}

\usepackage{pgfplots}
\usepackage{pgfplotstable}
\pgfplotsset{compat=1.14} 

\pgfplotsset{
	/pgf/declare function={
		Floor(\x) = round(\x-0.49);
	},
	show sum on top/.style={
		/pgfplots/scatter/@post marker code/.append code={%
			\path let \p1=($(normalized axis cs:%
			\pgfkeysvalueof{/data point/x},%
			\pgfkeysvalueof{/data point/y})%
			-(normalized axis cs:\pgfkeysvalueof{/data point/x},0)$)
			in node[
			at={(normalized axis cs:%
				\pgfkeysvalueof{/data point/x},%
				\pgfkeysvalueof{/data point/y})%
			},
			anchor={-90*sign(\y1)},yshift={sign(\y1)*2pt}
			]
			{\pgfmathprintnumber{\pgfkeysvalueof{/data point/y}}};
		},
	}
}

\usepackage{ragged2e}
\usepackage{amsmath}
\usepackage{makecell}

\newboolean{showcomments}
\setboolean{showcomments}{true}
\ifthenelse{\boolean{showcomments}}
{ \newcommand{\mynote}[2]{
		\fbox{\bfseries\sffamily\scriptsize#1}
		{\small$\blacktriangleright$\textsf{\textcolor{red}{{\em #2}\bf }}$\blacktriangleleft$}\GenericWarning{}{LaTeX Warning: TODO: #2}}}
{ \newcommand{\mynote}[2]{}}

\usepackage{colortbl}
\definecolor{pgreen}{RGB}{5,205,107}
\definecolor{pbrown}{RGB}{165,118,31}
\definecolor{pblue}{RGB}{2,154,223}
\definecolor{porange}{RGB}{242,133,34}
\definecolor{pred}{RGB}{255,33,93}
\definecolor{pgrey}{RGB}{159,177,186}

\usepackage{listings}
\usepackage{tikz}
\usepackage{xspace}
\usepackage{listings}
\lstdefinelanguage{Pom}{
	morekeywords={<dependency>,</dependency>,<groupId>,</groupId>,<artifactId>,</artifactId>,<version>,</version>,<scope>,</scope>,<plugin>,</plugin>,<executions>,</executions>,<execution>,</execution>,<goal>,</goal>,<goals>,</goals>,<configuration>,</configuration>,<entryClass>,</entryClass>,<entryMethod>,</entryMethod>,<entryParameters>,</entryParameters>},
	otherkeywords={<dependency>,</dependency>,<groupId>,</groupId>,<artifactId>,</artifactId>,<version>,</version>,<scope>,</scope>,<plugin>,</plugin>,<executions>,</executions>,<execution>,</execution>,<goal>,</goal>,<goals>,</goals>,<configuration>,</configuration>,<entryClass>,</entryClass>,<entryMethod>,</entryMethod>,<entryParameters>,</entryParameters>}
}

\lstdefinestyle{Pom}{
	language={Pom}, basicstyle=\ttfamily\normalsize
}

\mdfdefinestyle{mpdframe}{
	frametitlebackgroundcolor   =black!15,
	frametitlerule              =true,
	roundcorner                 =5pt,
	middlelinewidth             =0.8pt,
	innermargin                 =0.2cm,
	outermargin                 =0.2cm,
	innerleftmargin             =0.2cm,
	innerrightmargin            =0.2cm,
	innertopmargin              =0.2cm,
	innerbottommargin           =0.2cm
}

\usepackage{pifont}
\usepackage{color}
\usepackage{tikz}
\usepackage{xspace}
\usepackage{listings}
\makeatletter
\newenvironment{btHighlight}[1][]
{\begingroup\tikzset{bt@Highlight@par/.style={#1}}\begin{lrbox}{\@tempboxa}}
	{\end{lrbox}\bt@HL@box[bt@Highlight@par]{\@tempboxa}\endgroup}
\newcommand\btHL[1][]{%
	\begin{btHighlight}[#1]\bgroup\aftergroup\bt@HL@endenv%
	}
	\def\bt@HL@endenv{%
	\end{btHighlight}%
	\egroup
}
\newcommand{\bt@HL@box}[2][]{%
	\tikz[#1]{%
		\pgfpathrectangle{\pgfpoint{1pt}{0pt}}{\pgfpoint{\wd #2}{\ht #2}}%
		\pgfusepath{use as bounding box}%
		\node[anchor=base west, fill=orange!30,outer sep=0pt,inner xsep=0pt, inner ysep=0pt, minimum height=\ht\strutbox+1pt,#1]{\raisebox{1pt}{\strut}\strut\usebox{#2}};
	}%
}
\lstdefinestyle{Java}{
	language={Java}, basicstyle=\ttfamily\scriptsize, 
	moredelim=**[is][{\btHL[fill=red!17,thin]}]{`}{`},
	moredelim=**[is][{\btHL[fill=green!17,thin]}]{£}{£},
	moredelim=**[is][{\btHL[fill=yellow!17,thin]}]{~}{~},
	moredelim=**[is][{\color{green!17}\btHL[fill=green!17,thin]}]{¤}{¤},
	moredelim=**[is][{\color{red!17}\btHL[fill=red!17,thin]}]{µ}{µ},
}

\usetikzlibrary{calc}
\usetikzlibrary{decorations.pathmorphing}

\input{er.tex}  
\input{abbrev}
\usepackage{fp}
\newcommand{\ShowAbsoluteNumber}[1]{%
\ifnum #1<10%
{\hspace*{0pt}#1}%
\else%
\ifnum #1<100%
{\hspace*{0pt}#1}%
\else%
\ifnum #1<1000%
{\hspace*{0pt}#1}%
\else%
{\numprint{#1}}%
\fi%
\fi%
\fi%
}

\newcommand{\ShowPercentage}[2]{%
\FPeval\percentage{round(#1/#2*100,0)}%
\FPeval\percentageOneDecimal{round(#1/#2*100,1)}%
\ifnum \percentage=0%
{\np[\%]{\FPprint{percentageOneDecimal}}}%
\else%
\ifnum \percentage<10%
{\np[\%]{\FPprint{percentageOneDecimal}}}%
\else%
{\np[\%]{\FPprint{percentageOneDecimal}}}%
\fi%
\fi%
\xspace
}
\newlength\BARSIZE  
\newcommand{\inlinechart}[2]{%
\FPeval{\BLACKBARSIZE}{#1/#2}\textcolor{black!80}{\rule{\BLACKBARSIZE\BARSIZE}{1.6ex}}%
\FPeval{\BLACKBARSIZE}{1 - (#1/#2)}\textcolor{black!10}{\rule{\BLACKBARSIZE\BARSIZE}{1.6ex}}%
}

\newcommand*\ChartSmall[3][v]{%
\ifx q#1%
    \np{#2}/\np{#3}(\ShowPercentage{#2}{#3})\else%
\ifx p#1%
    \np{#2}(\ShowPercentage{#2}{#3})\else%
\ifx c#1%
    \inlinechart{#2}{#3}%
\else%
    \np{#2}%
    \ifx r#1%
        /\np{#3}%
    \fi%
    \hspace*{0.5ex}(\ShowPercentage{#2}{#3}) %
    \inlinechart{#2}{#3}%
    \xspace
\fi\fi\fi%
}

\newcommand{\ShowReduction}[2]{%
\FPeval\percentage{round(100-#1/#2*100,0)}%
\FPeval\percentageOneDecimal{round(100-#1/#2*100,1)}%
\ifnum \percentage=0%
{\np[\%]{\FPprint{percentageOneDecimal}}}%
\else%
\ifnum \percentage<10%
{\np[\%]{\FPprint{percentageOneDecimal}}}%
\else%
{\np[\%]{\FPprint{percentageOneDecimal}}}%
\fi%
\fi%
\xspace
}

\def\nbLib{94}
\def\nbLibStr{\np{\nbLib}}
\def\nbLibVersion{395}
\def\nbLibVersionStr{\np{\nbLibVersion}}
\def\nbLibCompiling{354}

\def\nbTest{713932}
\def\nbTestStr{\np{\nbTest}}
\def\nbTestClient{211116}
\def\nbTestClientStr{\np{\nbTestClient}}
\def\nbClientAll{2874}
\def\nbClientAllStr{\np{\nbClientAll}}

\def\medianCoverageLib{80.83}
\def\medianCoverageLibStr{\np[\%]{\medianCoverageLib}}

\def\medianCoverageClient{20.24}
\def\medianCoverageClientStr{\np[\%]{\medianCoverageClient}}
\def\nbLineLib{10831394}
\def\nbLineLibStr{\np{\nbLineLib}}
\def\nbLineClient{140910102}
\def\nbLineClientStr{\np{\nbLineClient}}
\def\executionTime{4 days, 8:39:09\xspace}
\def\debloatTime{1 day, 10:55:04\xspace}

\def\nbLibDebSuccessNum{302}
\def\nbLibNotCompile{41}
\def\nbLibCrash{29}
\def\nbLibTimeout{13}
\def\nbLibValidation{10}

\FPeval{nbDebloatError}{round(\nbLibVersion - \nbLibDebSuccessNum, 0)}

\def\nbLibDebSuccessNum{302}
\def\nbLibPassTestNum{211}
\def\nbLibExcludedForTest{61}
\def\nbLibTestFailing{30}
\def\nbUniqueTestNum{342835}

\def\nbFailingTest{1405}
\def\nbErrorTest{432}

\FPeval{nbLibExecTestNum}{round(\nbLibPassTestNum + \nbLibTestFailing, 0)}
\FPeval{nbPassingTestNum}{round(\nbUniqueTestNum - \nbFailingTest, 0)}
\FPeval{nbFailureTest}{round(\nbFailingTest - \nbErrorTest, 0)}

\def\nbLibWithDependencies{70}

\def\nbDependencies{187}
\def\nbClasses{103032}
\def\nbMethods{693703}
\def\nbBloatedDependencies{38}
\def\nbBloatedClasses{61929}
\def\nbBloatedMethods{411997}

\def\nbBloatedDepClasses{22758}

\def\nbBloatedDepMethods{172060}

\FPeval{nbLibWithoutDependencies}{round(\nbLibPassTestNum - \nbLibWithDependencies, 0)}
\FPeval{nbNonBloatedDependencies}{round(\nbDependencies - \nbBloatedDependencies, 0)}

\def\totalSize{1151513795}
\def\resourceSize{257982707}
\def\bytecodeSize{893531088}
\def\debloatedSize{596337540}
\def\debloatedMethodSize{13768627}

\FPeval{bloatedSize}{round(\debloatedSize + \debloatedMethodSize, 0)}
\FPeval{nonBloatedSize}{round(\bytecodeSize - \debloatedSize - \debloatedMethodSize, 0)}

\def\nbClient{1354}

\def\nbDynCoveringClient{281}
\def\nbStatCoveringClient{988}

\def\nbFailingTestClient{52}
\def\nbFailingTestProject{37}
\def\nbClientTest{44357}
\def\nbClientFailure{716}
\def\nbClientPassingInFailure{2280}
\def\nbClientDebloatError{38}

\def\nbTotalProject{147991}
\def\nbTotalProjectStr{\np{\nbTotalProject}}

\FPeval{nbCientDebloatNoCompError}{round(\nbStatCoveringClient - \nbClientDebloatError, 0)}
\FPeval{nbClientDebloatNoTestFail}{round(\nbDynCoveringClient - \nbFailingTestClient, 0)}

\FPeval{nbClientStaticallyPassing}{round(\nbStatCoveringClient-\nbClientDebloatError, 0)}
\FPeval{nbClientDynamicallyPassing}{round(\nbDynCoveringClient-\nbFailingTestClient, 0)}

\FPeval{nbTestsInFailure}{round(\nbClientFailure - \nbClientPassingInFailure, 0)}
\FPeval{nbFailingTestNum}{round(\nbUniqueTestNum - \nbPassingTestNum, 0)}
\FPeval{nbSuccessTestNum}{round(\nbPassingTestNum*100/\nbUniqueTestNum, 2)}
\def\nbLibPassTest{\np{\nbLibPassTestNum}}
\FPeval{nbLibFailTestNum}{round(\nbLibDebSuccessNum - \nbLibPassTestNum, 0)}

\newcommand{\ie}{i.e.\@\xspace}

\newcommand{\eg}{e.g.\@\xspace}

\newcommand{\etal}{et al.\@\xspace}

\newcommand{\wrt}{w.r.t.\@\xspace}

\newcommand{\mc}{Maven Central\@\xspace}

\newcommand\jdbl{{\textsf{JDBL}}\xspace}
\newcommand\jshrink{{\textsc{JShrink}}\xspace}

\newcommand\mv{{Maven}\xspace}
\newcommand\jar{\texttt{JAR}\xspace}

\makeatother

\AtBeginDocument{%
}

\usepackage{csvsimple}
\usepackage{longtable}
\usepackage{lscape}
\usepackage{array}
\csvstyle{csvResultStyle}{
	head to column names,
	tabular=|p{.18\linewidth}|p{.25\linewidth}|p{.26\linewidth}|p{.36\linewidth}|,
	table head=\hline \bfseries System call perturbation & \bfseries HTTP requests & \bfseries System calls & \bfseries Other observations \\ \hline,
	late after last line=\\\hline
}

\usepackage{array, etoolbox}
\newcounter{rowcount}
\usepackage{amsthm}
\theoremstyle{definition}

\usepackage[framemethod=tikz]{mdframed}
\mdfdefinestyle{mpdframe}{
	nobreak                     =true,
	backgroundcolor             =black!3,
	frametitlebackgroundcolor   =black!15,
	frametitlerule              =false,
	roundcorner                 =5pt,
	innermargin                 =0.2cm,
	innerleftmargin             =0.2cm,
	innerrightmargin            =0.2cm,
	innertopmargin              =0.2cm,
	innerbottommargin           =0.2cm,
	shadowsize                  =1pt
}

\AtBeginDocument{%
}

\AtBeginDocument{%
  \providecommand\BibTeX{{%
    \normalfont B\kern-0.5em{\scshape i\kern-0.25em b}\kern-0.8em\TeX}}}

\setcopyright{NONE}
\copyrightyear{X}
\acmYear{X}
\acmDOI{X}

\begin{document}
	
\title{Coverage-Based Debloating for Java Bytecode}

\author{C\'esar Soto-Valero}
\email{cesarsv@kth.se}
\orcid{0000-0003-0541-6411}
\affiliation{%
  \institution{KTH Royal Institute of Technology}
  \city{Stockholm}
  \country{Sweden}
}
\author{Thomas Durieux}
\email{tdurieux@kth.se}
\orcid{0000-0002-1996-6134}
\affiliation{%
	\institution{KTH Royal Institute of Technology}
	\city{Stockholm}
	\country{Sweden}
}
\author{Nicolas Harrand}
\email{nharrand@kth.se}
\orcid{0000-0002-2491-2771}
\affiliation{%
	\institution{KTH Royal Institute of Technology}
	\city{Stockholm}
	\country{Sweden}
}
\author{Benoit Baudry}
\email{baudry@kth.se}
\orcid{0000-0002-4015-4640}
\affiliation{%
	\institution{KTH Royal Institute of Technology}
	\city{Stockholm}
	\country{Sweden}
}

\makeatletter
\let\@authorsaddresses\@empty
\makeatother

\renewcommand{\shortauthors}{Soto-Valero, et al.}

\begin{abstract}
Software bloat is code that is packaged in an application but is actually  not necessary to run the application. 
The presence of software bloat is an issue for security, for performance, and for maintenance. In this paper, we introduce a novel technique for debloating, which we call coverage-based debloating. We implement the technique for one single language: Java bytecode. We leverage a combination of state-of-the-art Java bytecode coverage tools to precisely capture what parts of a project and its dependencies are used when running with a specific workload. Then, we automatically remove the parts that are not covered, in order to generate a debloated version of the project. We succeed to debloat \nbLibPassTestNum ~library versions from a dataset of \nbLibStr unique ~open-source Java libraries. The debloated versions are syntactically correct and preserve their original behavior according to the workload. Our results indicate that \np[\%]{68.3} of the libraries' bytecode and \ShowPercentage{\nbBloatedDependencies}{\nbDependencies} of their total dependencies can be removed through coverage-based debloating.

For the first time in the literature on software debloating, we assess the utility of debloated libraries with respect to client applications that reuse them. We select \np{\nbStatCoveringClient} client projects that either have a direct reference to the debloated library in their source code or which test suite covers at least one class of the libraries that we debloat. Our results show that \ShowPercentage{\nbClientDebloatNoTestFail}{\nbDynCoveringClient} of the clients, with at least one test that uses the library, successfully compile and pass their test suite when the original library is replaced by its debloated version.
\end{abstract}

\ccsdesc[500]{Software and its engineering~Software libraries and repositories}
\ccsdesc[500]{Software and its engineering~Software maintenance tools}
\ccsdesc[500]{Software and its engineering~Empirical software validation}

\keywords{software bloat, code coverage, program specialization, bytecode, software maintenance}

\settopmatter{printfolios=false}
\maketitle

\noindent\rule{\textwidth}{0.5pt}

\section{Introduction}\label{sec:introduction}

Software systems have a natural tendency to grow in size and complexity over time ~\cite{Wirth1995, Holzmann2015, Quach2017, Guo2021}.
A part of this growth comes with new features or bug fixes, while another part is due to useless code that accumulates over time. 
This phenomenon, known as software bloat, increases when building on top of software frameworks~\cite{Laperdrix2019,Koo2019,Rastogi2017}, as well as  with  code reuse~\cite{Zhong2019,Soto2020,Spinellis2021}.
Software debloating consists of automatically removing unnecessary code~\cite{Haas2020}. 
Automatic debloating poses several challenges: determine the location of the bloated parts ~\cite{Chen2017, Sharif2018, Qian2019}, and remove these parts while preserving the original behavior and providing useful features.
The problem of safely debloating real-world applications remains a long-standing software engineering endeavor today.

Most state-of-the-art debloating techniques target this problem using static analysis~\cite{Tip2002, Jiang2016, Sharif2018, Valero2021}, because it is scalable. Yet, the results lack precision in the presence of dynamic language features, which are prevalent in modern programming languages, and commonly used in practice~\cite{Li2018}.
Dynamic program analysis techniques outperform static approaches through the runtime collection of program usage information~\cite{Qian2019, Chen2017}.
However, capturing complete and precise dynamic usage information for debloating is challenging, especially at scale.

In this paper, we introduce coverage-based debloating for Java bytecode.
Our new approach, implemented in the \underline{\textsc{J}}ava \underline{\textsc{D}}e\underline{\textsc{bl}}oater (\jdbl) tool, handles the challenge of capturing precise dynamic usage by leveraging the industry-standard dynamic analysis techniques implemented in software coverage tools.
Based on this information, \jdbl automatically transforms the bytecode of the compiled project to remove the bloated code. 
\jdbl validates the syntactic correctness of the debloated project, as well as its behavior. To do so, it rebuilds the debloated project with the same configuration as the original and re-executes the test suite to check that the behavior of the original project is preserved.

The key technical contribution of our work consists in collecting accurate code coverage to minimize the risks of generating an ill-formed debloated software artifact (\ie, debloating and packaging a software project for reuse).
The loss of information in the compilation from source to bytecode, as well as the existence of software elements that are required but are not executed, are two essential challenges to precisely capture the code that can be safely removed.
Additionally, coverage tools do not handle third-party libraries, which is a primary source of software bloat~\cite{Ziegler2019,Agadakos2020, Soto2020}.
In \jdbl, we aggregate the coverage data collected by four coverage tools, to address those challenges. The tools  implement complementary, custom heuristics to cover the corner cases. \jdbl also extends the Maven build mechanism to collect coverage information for third-party libraries.


We evaluate \jdbl by debloating \nbLibPassTestNum ~versions from a dataset of \nbLibVersionStr ~versions of \nbLibStr unique ~open-source Java libraries.
This represents  a total of 10M+ lines of code analyzed, \np{\nbClasses} classes and \np{\nbDependencies} unique third-party dependencies.
We assess the effectiveness of our technique to preserve both syntactic correctness and the original behavior of these libraries. 
We quantify the impact of coverage-based debloating on the libraries' size at three  granularity levels: number of removed methods, classes, and dependencies.
\jdbl finds that \ShowPercentage{\nbBloatedClasses}{\nbClasses} of classes are bloated, and  \ShowPercentage{\nbBloatedDependencies}{\nbDependencies} of the third-party libraries can be completely removed. 
A comparison with JShrink \cite{Bruce2020}, the state-of-the-art tool for Java debloating, indicates that \jdbl achieves significantly larger reduction rates, while systematically preserving the original behavior.

For the first time in the literature of software debloating, we assess the usability of the debloated libraries with respect to actual usages, by building client programs that declare a dependency towards these libraries.
First, we check if the client program compiles correctly with the debloated library to assess binary compatibility. 
Then, we check if the program's test suite still passes.
We evaluate the utility of coverage-based debloating with respect to \np{\nbStatCoveringClient} ~programs that have at least one direct reference to the debloated library in their source code. 
For \ShowPercentage{\nbClientDebloatNoTestFail}{\nbDynCoveringClient} of programs which test suite covers at least one class of the library the test suite passes with the debloated libraries.

\jdbl is a Java debloating tool that combines diverse coverage data sources with bytecode removal transformations. It  validates the debloating results throughout the whole software build pipeline.
Unlike existing Java debloating techniques~\cite{Bruce2020,Jiang2016, Kalhauge2019, Tip2002, Soto2020}, our approach exploits the diversity of bytecode coverage tools to collect complete coverage information through the whole dependency tree.
The complete automation of the debloating procedure and our more reliable approach for collecting usage information allows us to evaluate \jdbl on the largest debloating dataset up to date.
Moreover, this is the first work in the debloating literature that assesses the utility of the debloated libraries with respect to their clients.
In summary, the contributions of this paper are the following:
\begin{itemize}[noitemsep]
	\item A practical, automated bytecode debloating approach for Java artifacts based on the collection of complete coverage information from multiple sources.
	\item An open-source tool, \jdbl, which executes throughout the \mv build pipeline and automatically generates debloated versions of Java artifacts.
	\item The largest empirical study on software debloating performed with \nbLibPassTestNum ~debloated libraries, and investigated code reduction at three granularity levels.
	\item The first assessment of the impact of debloated third-party libraries on their clients, with \np{\nbStatCoveringClient} clients of the libraries  that \jdbl successfully debloats.
\end{itemize}


\section{Motivating example}\label{sec:background}

\begin{figure}[t]
	\centering
	\input{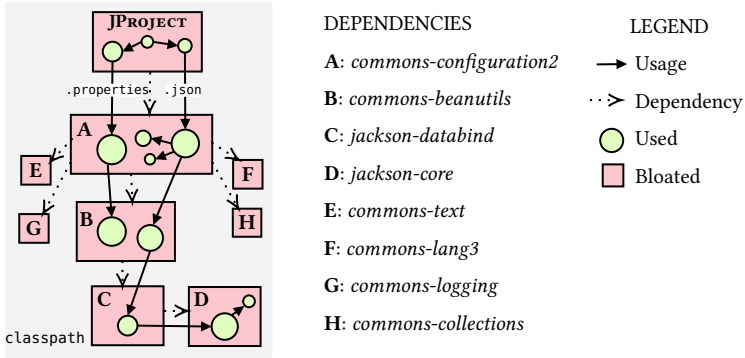}
	\caption{Typical code reuse scenario in the Java ecosystem. The Java project, \textsc{JProject}, uses functionalities provided by the library \textit{commons-configuration2}, which has $7$ dependencies. Rectangles, in red, represent Java artifacts. Circles  inside artifacts, in green, represent API members used by \textsc{JProject}.}
\label{fig:example_graph}
\vspace{-0.2cm}
\end{figure}

In this section, we illustrate the impact of software bloat in the context of a Java application with dependencies.
\autoref{fig:example_graph} shows the dependency tree of a typical Java project.
\textsc{JProject} implements a set of features and reuses functionalities provided by third-party dependencies.
To illustrate the notion of software bloat, we focus on one specific functionality that \textsc{JProject} reuses: parsing a configuration file located in the file system, provided by the \textit{commons-configuration2} library.\footnote{\scriptsize{\url{https://commons.apache.org/proper/commons-configuration2}}}

In our example, \textsc{JProject} uses this library to read \texttt{properties} and \texttt{json} configuration files. 
However, \textit{commons-configuration2} supports additional file formats, which are not necessary for \textsc{JProject} to run correctly, \ie, they are considered as bloat.
Yet, all the classes of the library must be added in the \texttt{classpath} of \textsc{JProject}, as well as all the runtime dependencies of the library.
The green circles and red squared components in \autoref{fig:example_graph} highlight this phenomenon: only the API members in green are necessary for \textsc{JProject}. All the code that belongs to the components in red, which includes all the functionalities for parsing other types of files than \texttt{properties} and \texttt{json}, are bloated with respect to \textsc{JProject}.
This represents a considerable amount of bytecode from \textit{commons-\allowbreak{}configuration2} that is included in \textsc{JProject} but is not needed. 
In addition, the dependency towards \textit{commons-configuration2}, implies that \textsc{JProject} has to include the classes of a total of \np{7} transitive dependencies in its \texttt{classpath}.
Some classes in the dependency \textbf{B} are used to process the file formats used by \textsc{JProject}, and parsing \texttt{json} files requires functionalities from dependencies \textbf{C} and \textbf{D}. 
Notice that the classes in the dependencies \textbf{E}, \textbf{F}, \textbf{G}, and \textbf{H}, are not necessary for \textsc{JProject}.

This example illustrates the  characteristics of Java projects: they are composed of a main module and import third-party dependencies. All the code of the main module, the dependencies, and the transitive dependencies is packaged in the project's \jar. Also, the existence of disjoint execution paths makes Java projects susceptible to include unnecessary functionalities from third-party libraries. 

In this paper, we focus on debloating functionalities from compiled Java projects and their dependencies. 
This involves the detection and removal of the reachable bytecode instructions that do not provide any functionalities to the project at runtime, both in the project's own classes and in the classes of its dependencies.
The objective of this bytecode transformation is to reduce the size of the project while still providing the same functionalities to its clients.

The main challenge for software debloating is to obtain precise usage information of the application and identify which parts can be safely removed.
In the next section, we describe our approach to overcome these challenges using code coverage. 
We motivate our approach and introduce the technical challenges. 
Then, we present the details of our technique.
\vspace{0.5cm}

\section{Coverage-Based Debloating}\label{sec:third}

Coverage-based debloating processes two inputs: a Java project, and coverage information collected when running a specific workload on the project. Our debloating technique removes the bytecode constructs that are not necessary to run the workload correctly. It produces a valid compiled Java project as output.
The debloated artifact is executable and has the same behavior as the original, \wrt the workload.

\begin{definition}
\textbf{Coverage-based debloating: }
Let $ \mathcal{P} $ be a program that contains a set of instructions $ \mathcal{S_P} $ and a workload that exercises a set $ \mathcal{F_P} $ of instructions, where $\mathcal{F_P} \subseteq \mathcal{S_P}$.
The coverage-based debloating technique transforms $ \mathcal{P} $ into a syntactically correct program $ \mathcal{P'} $, where $|\mathcal{S}_{P'}| \leq |\mathcal{S}_P|$ and $ \mathcal{P'} $ preserves the same behavior as $ \mathcal{P} $ when executing the workload.
\end{definition}

The collection of accurate coverage information is a critical task for coverage-based debloating.
In the following section, we discuss some key challenges and limitations of current techniques to collect complete Java bytecode coverage information.
Then, we introduce the solutions that we implement to address these technical challenges, which are part of our contributions.

\subsection{Challenges of Collecting Accurate and Complete Coverage for Debloating}\label{sec:complete-trace}

Java has a rich ecosystem of tools and algorithms to collect code coverage reports. 
These tools, which rely on bytecode transformations~\cite{Yang2009}, perform the following three key steps: 
(i) the bytecode is enriched with probes at particular locations of the program's control flow, depending on the granularity level of the coverage;
(ii) the instrumented bytecode is executed in order to collect the information on which probes are activated at runtime;
(iii) the activated regions of the bytecode are mapped with the source code, and a coverage report is given to the user.

Existing code coverage techniques are implemented in mature, robust, and scalable tools, which can serve as the foundation for coverage-based debloating. 
State-of-the-art tools for this purpose include JaCoCo,\footnote{\scriptsize{\url{https://www.eclemma.org/jacoco}}} JCov,\footnote{\scriptsize{\url{https://github.com/openjdk/jcov}}} and Clover.\footnote{\scriptsize{\url{https://openclover.org}}}
Yet, all of them have two essential limitations when used for debloating. 
First, different instrumentation strategies do not handle specific corner cases, while capturing the program's execution~\cite{Li2013}. 
For example, JaCoCo does not generate a complete coverage report for fields, methods that contain only one statement that triggers an exception, and the compiler-generated methods and classes for Java enumerations.
Second, by default, these tools collect coverage only for the bytecode of a compiled project and do not instrument the bytecode of third-party libraries.
In the following, we discuss the corner cases for accurate coverage in detail.
In \autoref{sec:solutions}, we present our approach to address corner cases and collect coverage information across the whole dependency tree.

Collecting code coverage involves several challenges related to source code compilation and bytecode instrumentation. 
First, the bytecode instrumentation must be safe and efficient, \ie, it must not alter the functional behavior of the application and have a limited runtime overhead. 
Second, the instrumentation must generate a coverage report that is complete, \ie, all the bytecode that is necessary to execute the workload should be reported as covered. 
This latter challenge is the most critical for coverage-based debloating: a single class missed in the report means that a necessary piece of bytecode will be removed, leading to an incorrect debloated application.

Three  factors affect the completeness of the coverage. 
First, no code coverage tool currently captures the coverage information across the whole dependency tree of a Java project.
This limits the effect of debloating based on code coverage to the project's sources only. 
Second, different tools have various instrumentation strategies to handle the variety of existing bytecode constructs~\cite{Horvath2019}. 
Consequently, these tools provide different reports for the same build setup.
Third, the Java compiler transforms the bytecode, causing information gaps between source and bytecode, \eg, by inlining constants or creating synthetic API members in certain situations~\cite{Lindholm2014, Tengeri2016}. 
In this case, it is not possible for coverage tools to collect information missing in the original bytecode. 
The following examples illustrate  five   challenges that we identified:

\textbf{Challenge \#1} \textit{Implicit Exceptions Thrown From Invoked Methods.}
\autoref{lst:exception} shows an example of an incorrect coverage report caused by a design limitation of JaCoCo. Both methods \texttt{m1} and \texttt{m2} are executed at runtime and both should be reported as covered. Yet,  \texttt{m1} (lines \ref{lst:m1-init} to \ref{lst:m1-end}) is missed by JaCoCo, while it is clear that, if we remove it, the \texttt{test} in class \texttt{FooTest} fails (lines \ref{lst:m2-test-init} to \ref{lst:m2-test-end}). This is because the JaCoCo probe insertion strategy does not consider implicit exceptions thrown from invoked methods.\footnote{\scriptsize{\url{https://www.eclemma.org/jacoco/trunk/doc/flow.html}}} These exceptions are subclasses of the classes \texttt{RuntimeException} and \texttt{Error}, and are expected to be thrown by the JVM itself at runtime. If the control flow between two probes is interrupted by an exception not explicitly created with a \texttt{throw} statement, all the instructions in between are missed by JaCoCo due to the non-existence of an instrumentation probe on the exit point of the method. 
In conclusion, JaCoCo misses one corner case for coverage: methods with a single-line invocation to other methods that throw exceptions.

\textbf{Challenge \#2} \textit{Implicit Methods in Enumerated Types.} \autoref{lst:enum} shows an example of incorrect coverage due to the inability of JaCoCo to account for implicit methods in enumerated types. \texttt{FooEnum} is a Java enumerated type declaring the string constant \texttt{MAGIC} with the value \texttt{"forty two"} (line \ref{lst:fortytwo}). The \texttt{test} method in the class \texttt{FooEnumTest} asserts the value of the constant in line \ref{lst:assert-enum}. However, the implicit method \texttt{valueOf}\footnote{\scriptsize{\url{https://docs.oracle.com/javase/7/docs/api/java/lang/Enum.html}}} in \texttt{FooEnum} is not covered according to JaCoCo. The reason is that, in Java, every enumerated type implicitly extends the class \texttt{java.lang.Enum}, which implements the methods \texttt{Enum.values()} and \texttt{Enum.\allowbreak{}valueOf()}. These methods are generated by the compiler, at compile-time. Therefore, they are not instrumented by coverage tools, which degrades the overall completeness of the produced coverage report.

\noindent\begin{minipage}{.45\textwidth}
\begin{lstlisting}[style=Java, escapechar=|, xleftmargin=3.5ex, caption={Example of an incomplete coverage report given by JaCoCo. The method \texttt{m1} is executed when running the method \texttt{test} in \texttt{FooTest}. However, this method is not considered as covered by JaCoCo.}, label={lst:exception},belowskip=-2\baselineskip]
public class Foo {
  `public void m1() {`|\label{lst:m1-init}|
   ` m2();`
  `}`|\label{lst:m1-end}|
  public void m2() {
    throw new IllegalArgumentException();
  }
}
 
public class FooTest {
  @Test(expected = IllegalArgumentException.class)|\label{lst:m2-test-init}|
  public void test() {
    Foo foo = new Foo();
    foo.m1();
  }|\label{lst:m2-test-end}|
}
\end{lstlisting}

\end{minipage}\hfill
\begin{minipage}{.5\textwidth}
\begin{lstlisting}[style=Java, escapechar=|, xleftmargin=3.5ex, caption={Example of an incomplete coverage result given by JaCoCo. The compiler-generated method \texttt{valueOf} in \texttt{FooEnum} is executed. However, this method is not instrumented and therefore is not reported as covered.}, label={lst:enum},belowskip=-2\baselineskip]
public enum FooEnum {
  MAGIC("forty two");|\label{lst:fortytwo}|
  public final String label;
  FooEnum(String label) { this.label = label; }
  `public static <T extends Enum<T>>` 
         `T valueOf(Class<T> enumType, String name)`
         `{...}`
}

public class FooEnumTest {
  @Test
  public void test() {
    assertEquals("forty two", 
          FooEnum.valueOf("MAGIC").label); |\label{lst:assert-enum}|
  }
}
\end{lstlisting}

\end{minipage}

\textbf{Challenge \#3} \textit{Java Compiler Optimizations.}
\autoref{lst:constant} illustrates an example that is incorrectly handled by all code coverage tools based on bytecode instrumentation. The variable \texttt{MAGIC}, initialized with a final static integer literal in line \ref{lst:magic-integer}, is used in the \texttt{FooTest} class as \texttt{Foo.MAGIC} (line \ref{lst:assert-magic-integer}). Therefore, the class \texttt{Foo} is necessary for the correct compilation and execution of the test method in the class \texttt{FooTest}. However, the class \texttt{Foo} is not detected as covered by JaCoCo or any other code coverage tool based on bytecode instrumentation. The cause is a bytecode optimization implemented in the \texttt{javac} compiler, which inlines constants at compilation time. This is shown in \autoref{lst:bytecode}, which is the bytecode generated after compiling the sources of the \texttt{FooTest} class from \autoref{lst:constant}. As we observe in lines \ref{lst:bipush0} to \ref{lst:bipush2}, the value of the constant \texttt{MAGIC} is directly substituted by its integer value, and hence the reference to the class \texttt{Foo} is lost during the compilation of the source code. Note that, if we remove the class \texttt{Foo}, the program will not compile correctly.

\noindent\begin{minipage}{.5\textwidth}
\begin{lstlisting}[style=Java, xleftmargin=3.5ex,escapechar=|, caption={Example of an inaccurate coverage report. The class \texttt{Foo} is not considered covered by any coverage tool, since the primitive constant \texttt{MAGIC} is inlined with its actual integer value by the Java compiler at compilation time.}, label={lst:constant},belowskip=-6\baselineskip]
`class Foo(){`
  `public static final int MAGIC = 42;`|\label{lst:magic-integer}|
`}`

public class FooTest {
  @Test
  public void test() {
    assertEquals(42, Foo.MAGIC);|\label{lst:assert-magic-integer}|
  }
}
\end{lstlisting}


\end{minipage}\hfill
\begin{minipage}{.45\textwidth}
\begin{lstlisting}[style=Java,  xleftmargin=3.5ex, escapechar=|, caption={Excerpt of the disassembled bytecode of \xspace~\autoref{lst:constant}. The Java compiler does not let any reference to the object \texttt{Foo} in the bytecode of the method \texttt{test} in class \texttt{FooTest}.}, label={lst:bytecode},belowskip=-5.75\baselineskip]
public class org.example.FooTest {
  public void test();
    Code:
       0: BIPUSH        42 |\label{lst:bipush0}|
       2: BIPUSH        42 |\label{lst:bipush2}|
       // Method junit/framework/TestCase.assertEquals:(II)V
       4: INVOKESTATIC  #3 
       7: RETURN
}
\end{lstlisting}

\end{minipage}\
\vspace{0.4cm}  

\textbf{Challenge \#4} \textit{Java Interfaces.} 
In \autoref{lst:interfaces}, the class \texttt{Foo} implements the method \texttt{doMagic} of the interface \texttt{Magic} (lines \ref{lst:init-magic-interface} to \ref{lst:end-magic-interface}).
This class will not compile correctly if its interface is removed.
However, JaCoCo does not instrument non-static methods in interfaces because they have no executable instructions.
Interfaces, exceptions, enumerations, and annotations are constructs of the Java language designed to facilitate software engineering tasks and most code coverage tools do not report them as covered.   

\textbf{Challenge \#5} \textit{Third-Party Dependencies.}
\autoref{lst:dependencies} presents an example of an used class from a third-party dependency that is not reported as covered by JaCoCo.
The class \texttt{Foo} uses the method \texttt{byteCountToDisplaySize} from the class \texttt{FileUtils} (line \ref{lst:dependency-usage}).
\texttt{FileUtils} is provided by the third-party dependency \textit{commons-io} and imported in line \ref{lst:dependency-import}.
However, when executing JaCoCo, the classes from this third-party are not instrumented.
This happens because JaCoCo is designed to cover only the project's code.

\vspace{0.4cm}

\noindent\begin{minipage}{.5\textwidth}
\begin{lstlisting}[style=Java, escapechar=|, xleftmargin=3.5ex, caption={Example of an incomplete coverage result given by JaCoCo. The interface \texttt{Magic} implemented by class \texttt{Foo} is necessary for the compilation of the class but it is not covered.}, label={lst:interfaces},belowskip=-2\baselineskip]
`public interface Magic {`|\label{lst:init-magic-interface}|
  `int doMagic();`
`}`|\label{lst:end-magic-interface}|

public class Foo implements Magic {
  @Override
  public int doMagic() {
    return 42;
  }
}

public class FooTest {
  @Test
  public void test() {
    Foo foo = new Foo();
    assertEquals(42, foo.doMagic());
  }
}
\end{lstlisting}

\end{minipage}\hfill
\begin{minipage}{.45\textwidth}
\begin{lstlisting}[style=Java, escapechar=|, xleftmargin=3.5ex, caption={Example of an incomplete coverage result given by JaCoCo. The class \texttt{FileUtils} in the dependency \textit{commons-io} is used but it is not covered.}, label={lst:dependencies},belowskip=-2\baselineskip]
`import org.apache.commons.io.FileUtils;`|\label{lst:dependency-import}|

public class Foo {
  public String showFileSize(long fileSize) {
    return FileUtils|\label{lst:dependency-usage}|
          .byteCountToDisplaySize(fileSize);
  }
}

public class FooTest {
  @Test
  public void test() {
    long fileSize = 50000;
    Foo foo = new Foo();
    assertEquals("48 KB", 
          foo.showFileSize(fileSize));
  }
}
\end{lstlisting}

\end{minipage}

\vspace{0.5cm}
\subsection{Addressing Coverage Challenges for Debloating}\label{sec:solutions}

This section describes our approach to tackle the bytecode coverage challenges presented in the previous section. 
The goal is to consolidate coverage information that can be used for debloating.

\subsubsection{Aggregating Coverage Reports}\label{sec:aggr_exec_trac}

We address the bytecode tracing challenges by aggregating the coverage reports produced by diverse coverage tools. 
The baseline coverage report is collected with JaCoCo. 
Then we consolidate this information as follows.

To  handle the case of implicit exceptions, illustrated in \autoref{lst:exception}, we develop Yajta,\footnote{\scriptsize{\url{https://github.com/castor-software/yajta}}} a customized tracing agent for Java. 
Yajta adds a probe at the beginning of the methods, including the default constructor. 
Yajta is based on  Javassist\footnote{\scriptsize{\url{https://www.javassist.org}}} for bytecode instrumentation. 
To handle compiler-generated methods, illustrated in \autoref{lst:enum}, we include the reports of JCov. This pure Java implementation of code coverage is officially maintained by Oracle and used for measuring coverage in the Java platform (JDK). It maintains the version of Java which is currently under development and supports the processing of large volumes of heterogeneous workloads.

We leverage the JVM class loader to obtain the list of classes that are loaded dynamically and lead to errors discussed in \autoref{lst:constant}.
The JVM dynamically links classes before executing them. The \texttt{-verbose:class} option of the JVM enables logging of class loading and unloading at runtime. 

\subsubsection{Keep All Necessary Bytecode That Cannot Be Covered}

The Java language contains specific constructs designed to achieve programming abstractions, \eg, interfaces, exceptions, enumerations, and annotations.
These elements do not execute any program logic and cannot be instantiated.
Therefore, they cannot be covered at runtime, and pure dynamic debloating  cannot determine if they are a source of bloat. 
Yet, they are necessary for compilation. 

To address this limitation, we always keep interfaces, enumeration types, exceptions, as well as static fields in the bytecode. 
This approach significantly improves the syntactic correctness of the debloated bytecode artifacts. Meanwhile, the impact on the size of the debloated code is minimal, due to the small size of such language constructs. 

\subsubsection{Capturing Coverage Across the Whole Dependency Tree}\label{sec:extension}

To effectively debloat a Java project, we need to analyze bytecode in the compiled project, as well as in its dependencies. To do so, we extend the coverage information provided by JaCoCo to the level of dependencies. This requires modifying the way JaCoCo interacts with \mv during the  build. 

We rely on the automated build infrastructure of \mv to compile the Java project and to resolve its dependencies. \mv provides dedicated plugins for fetching and storing all the dependencies of the project. Therefore, it is practical to rely on the \mv dependency management mechanisms, which are based on the  \textit{pom.xml} file that declares the direct dependencies of the project. 
These dependencies are  \jar files hosted in external repositories (\eg, \mv Central~\cite{Valero2019}).\footnote{\scriptsize{\url{https://repo.maven.apache.org/maven2}}}

Only dependencies in the runtime and compile \texttt{classpath} are packaged by \mv at the end of the build process.
Therefore, we focus on dependencies with these specific scopes.
Once the dependencies have been downloaded, we compile the Java sources and unpack all the bytecode of the project and its dependencies into a local directory. 
Then, probes are injected at the beginning and end of all Java bytecode methods of the classes in this directory. 
This code instrumentation is performed off-line, before the workload execution and coverage collection.
At runtime, the coverage tool is notified when the execution hits an injected probe. 
This way, our coverage-based approach captures the covered classes and methods in all dependencies. 

\subsection{Coverage-Based Debloating Procedure}\label{sec:approach}

In this section, we present the details of \jdbl, our end-to-end tool for automated coverage-based Java bytecode debloating.
\jdbl receives as input a Java project that builds correctly with \mv and a workload that exercises the project. 
\jdbl  outputs a debloated, packaged project that builds correctly and preserves the functionalities necessary to run that particular workload. 
The debloating procedure consists of three main phases. The coverage collection phase gathers usage information based on dynamic analysis. The bytecode removal phase modifies the bytecode of the artifact, based on coverage. The artifact validation phase assesses the correctness of the debloated artifact.

\autoref{algo:dynamic_debloat} details the three subroutines, corresponding to each debloating phase. 
In the following subsections, we describe these phases in more detail. 

\subsubsection{\ding{202} Coverage Collection.}

\jdbl collects a set of coverage reports that capture the set of dependencies, classes, and methods  actually used during the execution of the Java project. The coverage collection phase receives two inputs: a compilable set of Java sources, and a workload, \ie, a collection of entry-points and resources necessary to execute the compiled sources. The workload can be a set of test cases or a reproducible production workload. The coverage collection phase outputs the original, unmodified, bytecode and a set of coverage reports that account for the minimal set of classes and methods required to execute the workload. 

Lines \ref{line:algo:cp} to \ref{line:algo:end-trace} in \autoref{algo:dynamic_debloat} show this procedure.
It starts with the compilation of the input project $ \mathcal{P} $, resolving all its direct and transitive dependencies $ \mathcal{D} $, and adding the bytecode to the \texttt{classpath} $ CP $ of the project (line~\ref{line:algo:cp}). 
Then, the whole bytecode contained in $ CP $ (line~\ref{line:algo:inst}) is instrumented, and a data store is initialized to collect the classes and methods used when executing the workload $ \mathcal{W} $ (line~\ref{line:algo:usg}).
\jdbl executes  the instrumented bytecode with $ \mathcal{W} $, and the classes and methods used are saved (lines~\ref{line:algo:add-usg} and \ref{line:algo:end-trace}).
\jdbl considers $ \mathcal{W} $ to be the complete test suite of a Maven project, where each $w \in \mathcal{W}$ is an individual unit test executed by Maven.

\begin{algorithm}[t]
	\SetNoFillComment
	\KwData{A correct program $ \mathcal{P} $ that contains a set of source files $ \mathcal{S} $, and declares a set of dependencies $ \mathcal{D} $.} 
	\KwData{ A workload $ \mathcal{W} $ that exercises at least one functionality in $ \mathcal{P} $.}
	\KwResult{A correct version of $ \mathcal{P} $, called $ \mathcal{P'} $, which is smaller than $ \mathcal{P} $ and contains the necessary code to execute $ \mathcal{W} $ and obtain the same results as with $ \mathcal{P} $.}
	\tcp{\ding{202} \textit{Coverage collection phase}}   
	$CP$ $\leftarrow$ \textit{compileSources}($ \mathcal{S} $, $ \mathcal{P} $) $ \cup $ \textit{getDependencies}($ \mathcal{D} $, $ \mathcal{P} $)\; \label{line:algo:cp}
	$INST$ $\leftarrow$ \textit{instrument}($CP$)\; \label{line:algo:inst}
	$USG$ $\leftarrow$ $\emptyset$\; \label{line:algo:usg}
	\label{line:algo:resolve}
	\ForEach{$w \in \mathcal{W} $}{ \label{line:algo:init-trace}
		\textit{execute}$(w, INST)$\; \label{line:algo:exec-w}
		\ForEach{ $class\in INST$}{
			\If{\textit{isExecuted}$(class)$}{ 
				$USG$ $\leftarrow$ $ addKey(class, USG) $\; \label{line:algo:add-usg}
    			\ForEach{ $method \in class$}{
    				\If{\textit{isExecuted}$(method)$}{ 
    					$USG$ $\leftarrow$ $ addVal(method, class, USG) $\;\label{line:algo:end-trace}
    				}	
			    }
			}    
		}
	}
	\tcp{\ding{203} \textit{Bytecode removal phase}}
	\ForEach{$class \in CP$}{ \label{line:algo:init-remove}
		\eIf{$class \not\in keys(USG)$}{ 
			$CP$ $\leftarrow$ $CP \setminus class$\; \label{line:algo:remove-class}
		}{
			\ForEach{$method \in class$}{
				\If{$method \not\in values(class, USG)$}{
					$CP$ $\leftarrow$ $CP \setminus method$\; \label{line:algo:end-remove}
				}
			}
		}
	}
	\tcp{\ding{204} \textit{Artifact validation phase}}
	$OBS$ $\leftarrow$ \textit{execute}($\mathcal{W}$, $\mathcal{P}$)\; \label{line:algo:validate}
	\If{$ !buildSuccess(CP) \mid execute(\mathcal{W}, CP) \neq OBS $}{   \label{line:algo:check}
		\Return \textit{ALERT}\; 
	}
	$ \mathcal{P'}  $ $\leftarrow$ \textit{package}($CP$)\; \label{line:algo:package}
	\Return $ \mathcal{P'} $\; \label{line:algo:return}
	\caption{Coverage-based debloating procedure for a Java project.} 
	\label{algo:dynamic_debloat}
\end{algorithm}

\subsubsection{\ding{203} Bytecode Removal.}

The goal of the bytecode removal phase is to eliminate the methods, classes, and dependencies that are not used when running the project with the workload $ \mathcal{W} $. 
This procedure is based on the coverage information collected during the coverage collection phase.
The unused bytecode instructions are removed in two passes (lines~\ref{line:algo:init-remove} to \ref{line:algo:end-remove} in \autoref{algo:dynamic_debloat}). 
First, the unused class files and dependencies are directly removed from the \texttt{classpath} of the project (lines~\ref{line:algo:remove-class} and \ref{line:algo:end-remove}). 
Then, the procedure analyzes the bytecode of the classes that are covered.
When it encounters a method that is not covered, the body of the method is replaced to throw an \texttt{UsupportedOperationException}. 
We choose to throw an exception instead of removing the entire method to avoid JVM validation errors caused by the nonexistence of methods that are implementations of interfaces and abstract classes.


At the end of this phase, \jdbl has removed the bloated methods, classes, and dependencies. A method is considered bloated if it is not invoked while running the workload.
A class is considered bloated if it has not been instantiated or called via reflection and none of its fields or methods are used. 
A third-party dependency is considered bloated if none of its classes or methods are used when executing the project with a given workload.\footnote{In this work, we refer to \mv dependencies.}

\subsubsection{\ding{204} Artifact Validation.}

The goal of the artifact validation phase is to assess the syntactic and semantic correctness of the debloated artifact with respect to the workload provided as input. 
This is how we detect errors introduced by the bytecode removal, before packaging the debloated \jar.

To assess syntactic correctness, we verify the integrity of the  bytecode in the debloated version. 
This implies checking the validity of the bytecode that the JVM has to load at runtime, and also checking that no dependencies or other resources were incorrectly removed from the \texttt{classpath} of the \mv project. 
We reuse the \mv tool stack, which includes several validation checks at each step of the build process~\cite{Macho2021}. 
For example, \mv verifies the correctness of the \textit{pom.xml} file, and the integrity of the produced \jar at the last step of the build life-cycle.
To assess semantic correctness, we check that the debloated project executes correctly with the workload. 

\autoref{algo:dynamic_debloat} (lines~\ref{line:algo:validate} to \ref{line:algo:return}) details this last phase of coverage-based debloating. 
We run the original version of $ \mathcal{P} $ with the workload $ \mathcal{W} $, to collect the program's original outputs in the variable $ OBS $ (line~\ref{line:algo:validate}). 
Then, the algorithm performs two checks in line~\ref{line:algo:check}: 1) a syntactic check that passes if the build of the debloated program is successful; and 2) a behavioral check that passes if the debloated program produces the same output as $ \mathcal{P} $, with $ \mathcal{W} $.
In other words, it treats $ OBS $ as an oracle to check that the  debloated project preserves the behavior of $ \mathcal{P} $.
Finally, the debloated artifact is packaged and returned in line~\ref{line:algo:return}.

\subsubsection{Implementation Details}\label{sec:implementation}

The core implementation of \jdbl consists in the orchestration of mature code coverage tools and bytecode transformation techniques.
The coverage-based debloating process is integrated into the different \mv building phases. We focus on \mv as it is one of the most widely adopted build automation tools for Java artifacts.
It provides an open-source framework with the APIs required to resolve dependencies automatically and to orchestrate all the debloating phases during the project build.

\jdbl gathers direct and transitive dependencies by using the \texttt{maven-dependency}\footnote{\url{https://maven.apache.org/plugins/maven-dependency-plugin}} plugin with the \texttt{copy-dependencies} goal. This allows us to manipulate the project's \texttt{classpath} in order to extend code coverage tools at the level of dependencies, as explained in \autoref{sec:extension}. 
For bytecode analysis, the collection of non-removable classes, and the whole bytecode removal phase, we rely on ASM,\footnote{\url{https://asm.ow2.io}} a lightweight, and mature Java bytecode manipulation and analysis framework. 
The instrumentation of methods and the insertion of probes are performed by integrating JaCoCo, JCov, Yajta, and the JVM class loader within the \mv build pipeline, as described in \autoref{sec:aggr_exec_trac}. 

\jdbl is implemented as a multi-module \mv project with a total of 5\textit{K} lines of code written in Java. 
\jdbl is designed to debloat single-module Maven projects. 
It can be used as a \mv plugin that executes during the \textit{package} \mv phase. 
Thus, \jdbl is designed with usability in mind: it can be easily invoked within the \mv build life-cycle and executed automatically, no additional configuration or further intervention from the user is needed.
To use \jdbl, developers only need to add the \mv plugin within the build tags of the \textit{pom.xml} file.
The source code of \jdbl is publicly available on GitHub, with binaries published in \mc. More information on \jdbl is available at \url{https://github.com/castor-software/jdbl}.\\

\section{Empirical Study}\label{sec:methodology}  
In this section, we present our research questions, describe our experimental methodology, and the set of Java libraries utilized as study subjects.

\subsection{Research Questions} 

To evaluate our coverage-based debloating approach, we study its \textit{correctness}, \textit{effectiveness}, and \textit{impact}.
We assess the debloating results through four different validation layers: compilation and testing of the debloated Java libraries, and compilation and testing of their clients. 
Our study is guided by the following research questions:

\newcommand{\RQone}{\textit{To what extent can a generic, fully automated coverage-based debloating technique produce a debloated version of Java libraries?}}

\newcommand{\RQtwo}{\textit{To what extent do the debloated library versions preserve their original behavior \wrt the debloating workload?}}

\newcommand{\RQthree}{\textit{How much bytecode is removed in the compiled libraries and their dependencies?}}

\newcommand{\RQfour}{\textit{What is the impact of using the coverage-based debloating approach on the size of the packaged artifacts?}}

\newcommand{\RQfive}{\textit{How does coverage-based debloating compare with the state-of-the-art of Java debloating regarding the size of the packaged artifacts and behavior preservation?}}

\newcommand{\RQsix}{\textit{To what extent do the clients of debloated libraries compile successfully?}}

\newcommand{\RQseven}{\textit{To what extent do the clients behave correctly when using a debloated library?}}

{
\setlength\leftmargini{4em}
\begin{enumerate}[label=\textbf{RQ\arabic*}:, ref=RQ\arabic*]
	\item[\textbf{RQ1}:] \RQone 
	\item[\textbf{RQ2}:] \RQtwo
\end{enumerate}
}

\noindent RQ1 and RQ2 focus on assessing the \textit{correctness} of our approach.
In RQ1, we assess the ability of \jdbl at producing a valid debloated \jar for real-world Java projects. 
With RQ2, we analyze the behavioral correctness of the debloated artifacts.

{
\setlength\leftmargini{4em}
\begin{enumerate}[label=\textbf{RQ\arabic*}:, ref=RQ\arabic*]
	\item[\textbf{RQ3}:] \RQthree
	\item[\textbf{RQ4}:] \RQfour
	\item[\textbf{RQ5}:] \RQfive
\end{enumerate}
}

\noindent RQ3, RQ4, and RQ5 investigate the \textit{effectiveness} of our debloating procedure at producing a smaller artifact by removing the unnecessary bytecode. 
We measure this effectiveness with respect to the amount of debloated methods, classes, and dependencies, as well as with the reduction of the size of the bundled \jar files.

{
\setlength\leftmargini{4em}
\begin{enumerate}[label=\textbf{RQ\arabic*}:, ref=RQ\arabic*]
	\item[\textbf{RQ6}:] \RQsix
	\item[\textbf{RQ7}:] \RQseven
\end{enumerate}
}

\noindent In RQ6 and RQ7, we go one step further than any previous work on software debloating and investigate how coverage-based debloating of Java libraries impacts the clients of these libraries.
Our goal is to determine the ability of dynamic analysis via coverage at capturing the behaviors that are relevant for the users of the debloated libraries. 

\subsection{Data Collection}\label{sec:dataset} 

We have extracted a dataset of open-source \mv Java projects  from GitHub, which we use to answer our research questions.
We choose open-source projects because accessing closed-source software for research purposes is a difficult task. 
Moreover, the diversity of open-source software allows us to determine if our coverage-based debloating approach generalizes to a vast and rich ecosystem of Java projects.

The dataset is divided into two parts: a set of libraries, \ie, Java projects that are declared as a dependency by other Java projects, and a set of clients, \ie, Java projects that use the libraries from the first set.
The construction of this dataset is performed in $5$ steps:

\begin{enumerate}
\item We identify the \nbTotalProjectStr~Java projects on GitHub that have at least five stars. We use the number of stars as an indicator of interest~\cite{Borges2018}. 

\item We select the \ChartSmall[p]{34560}{\nbTotalProject} \mv projects that are single-module. We focus on single-module projects because they generate a single \jar. For this, we consider the projects that have a single \mv build configuration file (\ie, \textit{pom.xml}).

\item We ignore the projects that do not declare JUnit as a testing framework, and we exclude the projects that do not declare a fixed release, \eg, \texttt{LAST-RELEASE}, \texttt{SNAPSHOT}. We identify \ChartSmall[p]{155}{34560} libraries, and \ChartSmall[p]{25557}{34560} clients that use \np{2103} versions of the libraries.

\item We identify the commit associated with the version of the libraries, \eg, \texttt{commons-net:3.4} is defined in the commit SHA: \href{https://github.com/apache/commons-net/tree/74a2282b7e4c6905581f4f1b5a2ec412310cd5e7}{74a2282}. For this step, we download all the revisions of the \textit{pom.xml} files to identify the commit for which the release has been declared. We successfully identified the commit for \ChartSmall[q]{1026}{2103} versions of the libraries. \ChartSmall[q]{143}{155} libraries and \ChartSmall[q]{16964}{25557} clients are considered.

\item We execute three times the test suite of all the library versions and all clients, as a sanity check to filter out libraries with flaky tests. We keep the libraries and clients that have at least one test and have all the tests passing: \ChartSmall[q]{94}{143} libraries, \ChartSmall[q]{395}{1026} library versions, and \ChartSmall[q]{2874}{16964} clients passed this verification. From now on, we consider each library version as a unique library to improve the clarity of this paper.
\end{enumerate}

\begin{table*}[t]
\footnotesize
\centering
    \caption{Descriptive statistics of the dataset of libraries and their associated clients.}
    \label{tab:benchmark}
    \begin{tabularx}{.95\textwidth}{@{}l X rrrrrrr@{}}
    \toprule
    &                        & \textsc{Min} & \textsc{1st Qu.} & \textsc{Median} & \textsc{3rd Qu.} & \textsc{Max} & \textsc{Avg.} & \textsc{Total} \\
    \midrule
\multirow{4}{*}{\nbLibStr Libraries}

& \# Versions & \np{1} & \np{1} & \np{3.0} & \np{5.0} & \np{23.0} & \np{4.2} & \nbLibVersionStr \\

& \# Tests & \np{1} & \np{139.8} & \np{378.0} & \np{1108.2} & \np{24946} & \np{1830.6} & \nbTestStr \\
& \# LOC   & \np{132} & \np{5439.5} & \np{17935.5} & \np{47866.0} & \np{341429} & \np{35629.6} & \nbLineLibStr \\
& Total Class Coverage & \np[\%]{2.7} & \np[\%]{74.6} & \np[\%]{94.0} & \np[\%]{99.0} & \np[\%]{100.0} & \np[\%]{84.8} & N.A \\
\hline
\multirow{3}{*}{\nbClientAllStr Clients} 
& \# Tests & \np{1} & \np{4.5} & \np{20.0} & \np{74.0} & \np{11415} & \np{107.7} & \nbTestClientStr \\
& \# LOC   & \np{0} & \np{3130.0} & \np{9170.0} & \np{58990.0} & \np{4531710} & \np{72897.1} & \nbLineClientStr \\
& JaCoCo Coverage & \np[\%]{0.0} & \np[\%]{2.1} & \np[\%]{20.24} & \np[\%]{57.7} & \np[\%]{100.0} & \np[\%]{31.4} & N.A \\
\bottomrule
\end{tabularx}
\end{table*}


\autoref{tab:benchmark} summarizes the descriptive statistics of the dataset.
The total class coverage of the libraries is computed based on the aggregation of the coverage reports of the tools presented in \autoref{sec:aggr_exec_trac}.
The number of LOC and the coverage of the clients are computed with JaCoCo.
In total, our dataset includes \nbLibVersionStr Java libraries from \nbLibStr different repositories and \nbClientAllStr clients. 
The \nbLibVersionStr libraries include \nbTestStr test cases that cover \medianCoverageLibStr of the \nbLineLibStr LOC. 
One library in our dataset can generate fake Pokemons \cite{java-fake}. 
The clients have \nbTestClientStr test cases that cover \medianCoverageClientStr of the \nbLineClientStr LOC. 
The dataset is described in detail in Durieux \etal~\cite{Durieux2021}.

%

\subsection{Experimental Protocol}\label{sec:protocol}

In this section, we introduce the experimental protocol that we use to answer our research questions. 
The goal is to examine the ability of \jdbl to debloat Java projects configured to build with \mv.

For our experiments, we use the test suite of the projects as a workload. 
Test suites are widely available while obtaining a realistic workload for hundreds of libraries is extremely difficult.
Another motivation  is to integrate \jdbl  in the build process and deploy the debloated version, which can then be directly used by the clients.

We experiment  coverage-based debloating on \nbLibVersionStr different versions of \nbLibStr libraries. 
An original step in our experimental protocol consists of further validating the utility of the debloated libraries with respect to their clients. 
This way, we check if coverage-based debloating preserves the elements that are required to compile and successfully run the test suites of the clients.

\subsubsection{Coverage-Based Debloating Execution}\label{sec:jdbl_execution}

\begin{figure*}
\centering
\resizebox{\columnwidth}{!}{ \input{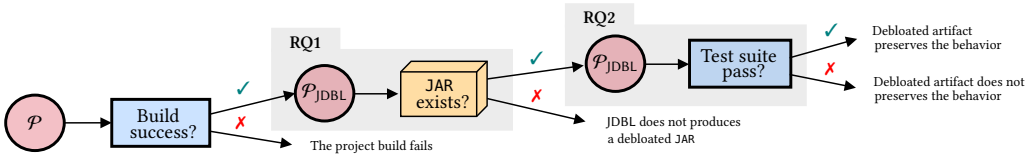}}
\caption{Pipeline of our experimental protocol to answer RQ1 and RQ2.}
\label{fig:rq12workflow}
\vspace{-0.2cm}
\end{figure*}

To run \jdbl at scale, we created an execution framework that automates the execution of our experimental pipeline. 
The framework orchestrates the execution of \jdbl and the collection of data to answer our research questions.
As \jdbl is implemented as a \mv plugin, most of the steps rely on the \mv build life-cycle.

The execution of \jdbl is composed of three main steps:
\begin{enumerate}[noitemsep]
\item \textit{Compile and test the original library.}
We build the original library (\ie, using \texttt{mvn package}) to ensure that it builds correctly and that all its test cases pass.
We configure the project to generate a \jar file  that contains all the binaries of the project.
This change of configuration may be in conflict with the original project build configuration and therefore fail in some scenarios.
At the end of the execution of the test suite, a fat \jar file is produced, which contains the bytecode of the library and all its dependencies.
We also store the reports concerning test execution and the corresponding logs.
The data produced during this step is used as a reference for further comparison with respect to the debloated version of the library, in RQ1 and RQ2.\looseness=-1

\item \textit{Configure the library to run} \jdbl.
The second step injects \jdbl as a plugin inside the \mv configuration (\textit{pom.xml}) and resets the configuration of the \texttt{maven-surefire-plugin} for our experiments.\footnote{\scriptsize{\url{https://maven.apache.org/surefire/maven-surefire-plugin}}}
This reset  ensures that its original configuration is not in conflict with the execution of the coverage collection phase of \jdbl.
A manual configuration of \jdbl could prevent this problem. Yet, we decided to standardize the execution for all the libraries   in order to scale up and automate the evaluation. 

\item \textit{Execute }\jdbl.
The third of our experiment framework executes \jdbl on the library, \ie, it runs \texttt{mvn package} with \jdbl configured in the \textit{pom.xml}.
At the end of this step, we collect the report generated by \jdbl with information about the debloated \jar (for RQ1, RQ3, and RQ4), the coverage report, and the test execution report (for RQ2).
\end{enumerate}

The execution was performed on a workstation running Ubuntu Server with an \texttt{i9-10900K} CPU (\np[cores]{16}) and \np[GB]{64} of RAM.
We set a maximum timeout constraint of 1:00:00 per project, which allows scaling up our experiments without an excessive debloating time.
It took \executionTime to execute the complete \jdbl experiment on our dataset, and \debloatTime to only debloat the libraries.
Each debloating execution is performed inside a Docker image in order to eliminate any potential side effects. 
The Docker image that we used during our experiment is available on DockerHub: \href{https://hub.docker.com/repository/docker/tdurieux/jdbl}{tdurieux/jdbl} which uses \jdbl commit SHA: \href{https://github.com/castor-software/jdbl/tree/c57396a5739e6ac3b0fa434342eb57b6f945914b}{c57396a}.
The execution framework is publicly available on GitHub \cite{jdbl-expe}, and the raw data obtained from the complete execution is available on Zenodo: \href{http://doi.org/10.5281/zenodo.3975515}{10.5281/zenodo.3975515}.
The \jdbl execution framework is composed of 3\textit{K} lines of Python code.\looseness=-1

\subsubsection{Debloating Correctness (RQ1 \& RQ2)}\label{sec:correctness_protocol}

To answer RQ1 and RQ2, we run \jdbl on each of the \nbLibVersionStr versions of \nbLibStr libraries.
RQ1 assesses the ability of \jdbl to produce a debloated \jar file, \ie, to successfully build the debloated \mv project. For RQ2, we analyze whether the test suite of the library has the same behavior before and after debloating. 

\autoref{fig:rq12workflow} illustrates the pipeline of RQ1 and RQ2.
First, we check that the library compiles correctly before the debloat.
If it does, then we verify if \jdbl has generated a \jar (RQ1).
If no \jar file is generated, then the debloating is considered as failed and the library is excluded for the rest of the evaluation.
The last step verifies that the test suite behaves the same before and after the bytecode removal phase.
This approach is consistent with previous works~\cite{Bruce2020,Ponta2021} in which existing tests are executed, and the results are used as a proxy for semantic preservation.

We compare the test execution reports produced during the first step of the \jdbl execution (see \autoref{sec:jdbl_execution}) and the test report generated during the verification step of \jdbl. 
We consider that the test suite has the same behavior on both versions if the number of executed tests is the same for both versions, and if the number of passing tests is also the same.
The number of executed tests might vary between the two versions because we modify the \texttt{maven-surefire-plugin} configuration to run as default in order to standardize and scale our experiments.
If the number of passing tests is not the same between the two reports, \jdbl is considered as having failed and the libraries are excluded for the rest of the evaluation.
We manually analyze the execution logs of the failing debloating executions to understand what happened.\looseness=-1

\subsubsection{Debloating Effectiveness (RQ3, RQ4, and RQ5)}\label{sec:effectiveness_protocol}

We assess the effectiveness of \jdbl regarding two different aspects.
The first aspect is related to code removal, checking the number of classes and methods that are debloated.
The second aspect is the size on disk that \jdbl allows saving by removing unnecessary parts of the libraries.

To answer RQ3, RQ4, and RQ5, we use the debloating reports of the original and debloated \jar files.
These reports contain the list of all the methods and classes of the libraries (including the dependencies), and if the element was debloated or not.
For RQ3, we compute the ratio of methods and classes that are debloated.
For RQ4, we extract the original and debloated \jar, and we compare the size in bytes of all the extracted files. 
To answer RQ5, we compare the bytecode size reduction and the test results after debloating with \jdbl and  with \jshrink. 
\jshrink is the most recent tool for debloating Java bytecode applications using dynamic analysis.
The source code of \jshrink is publicly available, and its debloating capabilities for a benchmark of Java projects are presented in its companion research paper~\cite{Bruce2020}.


For RQ3 and RQ4, we consider the \nbLibPassTest library versions that successfully pass the debloating correctness assessment.
We separate the \ChartSmall[q]{\nbLibWithoutDependencies}{\nbLibPassTestNum} libraries that do not have dependencies and the \ChartSmall[q]{\nbLibWithDependencies}{\nbLibPassTestNum} libraries that have at least one dependency.
We decided to do so because we observed that the libraries that have dependencies contain many more elements (bytecode and resources), which may negatively impact the analysis compared to libraries that do not have a dependency. For RQ5, we consider $17$ Java projects in the original benchmark used to evaluate \jshrink and compare \jdbl against the the debloating results reported in the \jshrink paper~\cite{Bruce2020}.

\subsubsection{Debloating Impact on Clients (RQ6 and RQ7)}\label{sec:rq5_protocol}

\begin{figure}
\centering
\subcaptionbox{RQ6 pipeline.\label{fig:rq5workflow}}{
	\resizebox{\columnwidth/2 + 3cm}{!}{\tikzset{every picture/.style={line width=0.75pt}} 

\begin{tikzpicture}[x=0.72pt,y=0.72pt,yscale=-1,xscale=1]

\draw [line width=0.75]    (107.58,1635.4) -- (132.18,1635.76) ;
\draw [shift={(135.18,1635.8)}, rotate = 180.83] [fill={rgb, 255:red, 0; green, 0; blue, 0 }  ][line width=0.08]  [draw opacity=0] (8.93,-4.29) -- (0,0) -- (8.93,4.29) -- cycle    ;
\draw  [fill={rgb, 255:red, 208; green, 2; blue, 27 }  ,fill opacity=0.25 ][line width=1.5]  (61.67,1635.84) .. controls (61.67,1623.93) and (71.99,1614.27) .. (84.73,1614.27) .. controls (97.46,1614.27) and (107.78,1623.93) .. (107.78,1635.84) .. controls (107.78,1647.75) and (97.46,1657.4) .. (84.73,1657.4) .. controls (71.99,1657.4) and (61.67,1647.75) .. (61.67,1635.84) -- cycle ;
\draw  [fill={rgb, 255:red, 50; green, 144; blue, 255 }  ,fill opacity=0.25 ][line width=1.5]  (136.03,1616.58) .. controls (136.03,1616.58) and (136.03,1616.58) .. (136.03,1616.58) -- (210.58,1616.58) .. controls (210.58,1616.58) and (210.58,1616.58) .. (210.58,1616.58) -- (210.58,1653.09) .. controls (210.58,1653.09) and (210.58,1653.09) .. (210.58,1653.09) -- (136.03,1653.09) .. controls (136.03,1653.09) and (136.03,1653.09) .. (136.03,1653.09) -- cycle ;
\draw [line width=0.75]    (345.03,1602.55) -- (397.4,1591.85) ;
\draw [shift={(400.34,1591.25)}, rotate = 528.45] [fill={rgb, 255:red, 0; green, 0; blue, 0 }  ][line width=0.08]  [draw opacity=0] (8.93,-4.29) -- (0,0) -- (8.93,4.29) -- cycle    ;
\draw [line width=0.75]    (345.03,1617.55) -- (398.61,1633.92) ;
\draw [shift={(401.48,1634.8)}, rotate = 196.99] [fill={rgb, 255:red, 0; green, 0; blue, 0 }  ][line width=0.08]  [draw opacity=0] (8.93,-4.29) -- (0,0) -- (8.93,4.29) -- cycle    ;
\draw  [fill={rgb, 255:red, 50; green, 144; blue, 255 }  ,fill opacity=0.25 ][line width=1.5]  (267.18,1592.58) .. controls (267.18,1592.58) and (267.18,1592.58) .. (267.18,1592.58) -- (344.34,1592.58) .. controls (344.34,1592.58) and (344.34,1592.58) .. (344.34,1592.58) -- (344.34,1629.09) .. controls (344.34,1629.09) and (344.34,1629.09) .. (344.34,1629.09) -- (267.18,1629.09) .. controls (267.18,1629.09) and (267.18,1629.09) .. (267.18,1629.09) -- cycle ;
\draw [line width=0.75]    (211.03,1626.55) -- (263.4,1615.85) ;
\draw [shift={(266.34,1615.25)}, rotate = 528.45] [fill={rgb, 255:red, 0; green, 0; blue, 0 }  ][line width=0.08]  [draw opacity=0] (8.93,-4.29) -- (0,0) -- (8.93,4.29) -- cycle    ;
\draw [line width=0.75]    (211.03,1641.55) -- (264.61,1657.92) ;
\draw [shift={(267.48,1658.8)}, rotate = 196.99] [fill={rgb, 255:red, 0; green, 0; blue, 0 }  ][line width=0.08]  [draw opacity=0] (8.93,-4.29) -- (0,0) -- (8.93,4.29) -- cycle    ;

\draw (84.73,1635.84) node   [align=left] {$\mathcal{C}_\mathcal{P}$};
\draw (173.92,1640.95) node   [align=left] {statically?};
\draw (172.92,1627.95) node   [align=left] {Is $\mathcal{P}$ used};
\draw (375.85,1585.67) node   [align=left] {{\small \textcolor{teal}{\textbf{\ding{51}}}}};
\draw (375.85,1614.67) node   [align=left] {{\small \textcolor{red}{\textbf{\ding{55}}}}};
\draw (307.92,1617.95) node   [align=left] {compiles?};
\draw (303.92,1604.95) node   [align=left] {Client};
\draw (327.85,1666.67) node   [align=left] {{\scriptsize use the library statically}};
\draw (317.85,1654.67) node   [align=left] {{\scriptsize The client does not }};
\draw (241.85,1609.67) node   [align=left] {{\small \textcolor{teal}{\textbf{\ding{51}}}}};
\draw (241.85,1638.67) node   [align=left] {{\small \textcolor{red}{\textbf{\ding{55}}}}};
\draw (476.85,1584.67) node   [align=left] {{\scriptsize The debloated library does not}};
\draw (465.85,1596.67) node   [align=left] {{\scriptsize breaks client compilation}};
\draw (456.85,1627.67) node   [align=left] {{\scriptsize The debloated library}};
\draw (465.85,1639.67) node   [align=left] {{\scriptsize breaks client compilation}};

\end{tikzpicture}}
}\\
\vspace{0.3cm}
\subcaptionbox{RQ7 pipeline.\label{fig:rq6workflow}}{
	\resizebox{\columnwidth/2 + 3cm}{!}{\tikzset{every picture/.style={line width=0.75pt}} 

\begin{tikzpicture}[x=0.73pt,y=0.73pt,yscale=-1,xscale=1]

\draw [line width=0.75]    (106.58,1779.4) -- (131.18,1779.76) ;
\draw [shift={(134.18,1779.8)}, rotate = 180.83] [fill={rgb, 255:red, 0; green, 0; blue, 0 }  ][line width=0.08]  [draw opacity=0] (8.93,-4.29) -- (0,0) -- (8.93,4.29) -- cycle    ;
\draw  [fill={rgb, 255:red, 208; green, 2; blue, 27 }  ,fill opacity=0.25 ][line width=1.5]  (60.67,1779.84) .. controls (60.67,1767.93) and (70.99,1758.27) .. (83.73,1758.27) .. controls (96.46,1758.27) and (106.78,1767.93) .. (106.78,1779.84) .. controls (106.78,1791.75) and (96.46,1801.4) .. (83.73,1801.4) .. controls (70.99,1801.4) and (60.67,1791.75) .. (60.67,1779.84) -- cycle ;
\draw  [fill={rgb, 255:red, 50; green, 144; blue, 255 }  ,fill opacity=0.25 ][line width=1.5]  (134.48,1760.58) .. controls (134.48,1760.58) and (134.48,1760.58) .. (134.48,1760.58) -- (228.13,1760.58) .. controls (228.13,1760.58) and (228.13,1760.58) .. (228.13,1760.58) -- (228.13,1797.09) .. controls (228.13,1797.09) and (228.13,1797.09) .. (228.13,1797.09) -- (134.48,1797.09) .. controls (134.48,1797.09) and (134.48,1797.09) .. (134.48,1797.09) -- cycle ;
\draw [line width=0.75]    (362.03,1746.55) -- (414.4,1735.85) ;
\draw [shift={(417.34,1735.25)}, rotate = 528.45] [fill={rgb, 255:red, 0; green, 0; blue, 0 }  ][line width=0.08]  [draw opacity=0] (8.93,-4.29) -- (0,0) -- (8.93,4.29) -- cycle    ;
\draw [line width=0.75]    (362.03,1761.55) -- (415.61,1777.92) ;
\draw [shift={(418.48,1778.8)}, rotate = 196.99] [fill={rgb, 255:red, 0; green, 0; blue, 0 }  ][line width=0.08]  [draw opacity=0] (8.93,-4.29) -- (0,0) -- (8.93,4.29) -- cycle    ;
\draw  [fill={rgb, 255:red, 50; green, 144; blue, 255 }  ,fill opacity=0.25 ][line width=1.5]  (284.18,1736.58) .. controls (284.18,1736.58) and (284.18,1736.58) .. (284.18,1736.58) -- (361.34,1736.58) .. controls (361.34,1736.58) and (361.34,1736.58) .. (361.34,1736.58) -- (361.34,1773.09) .. controls (361.34,1773.09) and (361.34,1773.09) .. (361.34,1773.09) -- (284.18,1773.09) .. controls (284.18,1773.09) and (284.18,1773.09) .. (284.18,1773.09) -- cycle ;
\draw [line width=0.75]    (229.03,1769.55) -- (281.4,1758.85) ;
\draw [shift={(284.34,1758.25)}, rotate = 528.45] [fill={rgb, 255:red, 0; green, 0; blue, 0 }  ][line width=0.08]  [draw opacity=0] (8.93,-4.29) -- (0,0) -- (8.93,4.29) -- cycle    ;
\draw [line width=0.75]    (229.03,1784.55) -- (282.61,1800.92) ;
\draw [shift={(285.48,1801.8)}, rotate = 196.99] [fill={rgb, 255:red, 0; green, 0; blue, 0 }  ][line width=0.08]  [draw opacity=0] (8.93,-4.29) -- (0,0) -- (8.93,4.29) -- cycle    ;

\draw (83.73,1779.84) node   [align=left] {$\mathcal{C}_\mathcal{P}$};
\draw (392.85,1729.67) node   [align=left] {{\small \textcolor{teal}{\textbf{\ding{51}}}}};
\draw (392.85,1758.67) node   [align=left] {{\small \textcolor{red}{\textbf{\ding{55}}}}};
\draw (324.92,1761.95) node   [align=left] {pass?};
\draw (320.92,1748.95) node   [align=left] {Test suite};
\draw (181.92,1784.95) node   [align=left] {dynamically?};
\draw (182.92,1771.95) node   [align=left] {Is $\mathcal{P}$ used};
\draw (335.85,1795.67) node   [align=left] {{\scriptsize The client does not }};
\draw (352.85,1807.67) node   [align=left] {{\scriptsize use the library dynamically}};
\draw (448.85,1740.67) node   [align=left] {{\scriptsize its behavior}};
\draw (466.85,1728.67) node   [align=left] {{\scriptsize The client preserves }};
\draw (473.85,1780.67) node   [align=left] {{\scriptsize preserves its behavior}};
\draw (465.85,1768.67) node   [align=left] {{\scriptsize The client does not}};
\draw (259.85,1752.67) node   [align=left] {{\small \textcolor{teal}{\textbf{\ding{51}}}}};
\draw (259.85,1781.67) node   [align=left] {{\small \textcolor{red}{\textbf{\ding{55}}}}};

\end{tikzpicture}}
}
\caption{Pipelines of our experimental protocol to answer RQ6 and RQ7.}
\label{fig:protocols}
\vspace{-0.3cm}
\end{figure}
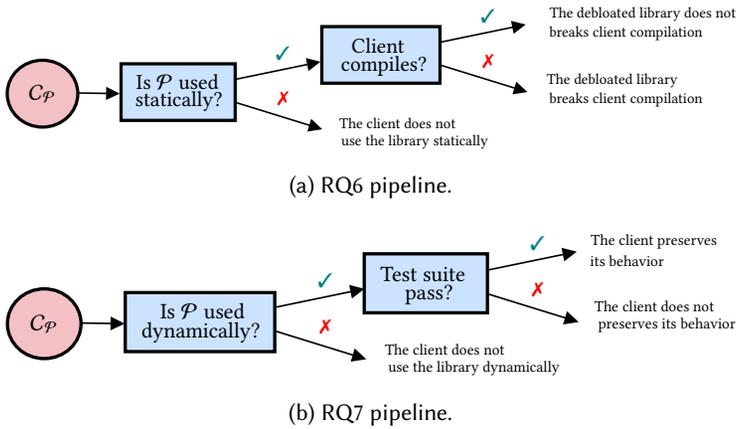

In the two final research questions, we analyze the impact of debloating Java libraries on their clients.
This analysis is relevant since we are debloating libraries that are mostly designed to be used by clients. 
This analysis also provides further information on the validity of this approach.
As far as we know, this is the first time that a software debloating technique is validated with the clients of the debloated artifacts.
We perform debloating validation from the clients' side at two layers: client's compilation and client's testing.

For RQ6, we verify that the clients still compile when the original library is replaced by its debloated version. 
We check that \jdbl does not remove classes or methods in libraries that are necessary for the compilation of their client.
\autoref{fig:rq5workflow} illustrates the pipeline for this research question. 
First, we check that the client $\mathcal{C}_\mathcal{P}$  uses the library statically in the source code. 
To do so, we analyze the source code of the clients.
If there is at least one element from the library present in the source code of a client, then we consider the library as statically used by the client. 

If the library is used, we inject the debloated library and build the client again. 
If the client successfully compiles, we conclude that \jdbl debloated the library while preserving the useful parts of the code that are required for compilation.

A debloated library stored on disk is of little use compared to a debloated library that provides the behavior expected by its clients. Therefore, with RQ7 we wish to determine if \jdbl preserves the functionalities that are necessary for the clients. \autoref{fig:rq6workflow} illustrates the pipeline for this research question. First, we execute the test suite of the client $\mathcal{C}_\mathcal{P}$  with the original version of the library. We check that the library is covered by at least one test of the client. 
If this is true, we replace the library with the debloated version and execute the test suite again.
If the test suite behaves the same as with the original library, we conclude that \jdbl is able to preserve the functionalities that are relevant for the clients.

To ensure the validity of this protocol, we perform additional checks on the clients.
All the clients have to use at least one of the \nbLibPassTest debloated libraries. 
We only consider the \ChartSmall[q]{\nbStatCoveringClient}{\nbClient} clients that either have a direct reference to the debloated library in their source code or which test suite covers at least one class of the library (static or dynamic usage).
The \np{\nbStatCoveringClient} clients that statically use the library serve as the study subjects to answer RQ6. 
The \ChartSmall[q]{\nbDynCoveringClient}{\nbStatCoveringClient} clients that have at least a test that reaches the debloated library serve as the study subjects to answer RQ7.

\section{Results}\label{sec:results}

We present our experimental results on the correctness, effectiveness, and impact of  coverage-based debloating for automatically removing unnecessary bytecode from Java projects.

\subsection{Debloating Correctness (RQ1 and RQ2)}

In this section, we report on the successes and failures of \jdbl to produce a correct debloated version of Java libraries.

\subsubsection{RQ1. \RQone}

In the first research question, we evaluate the ability of \jdbl at performing automatic coverage-based debloating for the \nbLibVersionStr libraries in our initial dataset. 
Here, we consider the debloating procedure to be successful if \jdbl produces a valid debloated \jar file for a library. 
To reach this successful state, the project to be debloated must go through all the build phases of the \mv build life-cycle, \ie, compilation, testing, and packaging, according to the protocol described in \autoref{sec:correctness_protocol}.

\begin{figure}
\centering
\resizebox{\columnwidth/2 + 3cm}{!}{\input{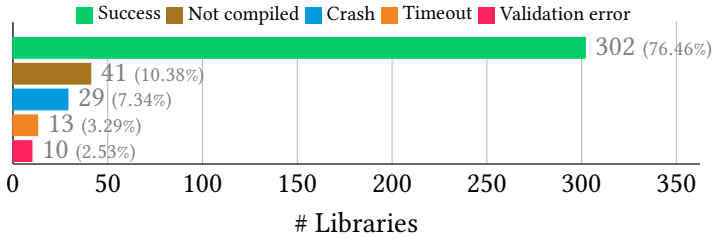}}
\caption{Number of libraries for which \jdbl succeeds or fails to produce a debloated \jar file.}
\label{fig:debloat_effectiveness_stacked_barplot}
\end{figure}

\autoref{fig:debloat_effectiveness_stacked_barplot} shows a bar plot of the number of successfully debloated libraries. It also displays the number of cases where \jdbl does not produce a debloated \jar file, due to failures in the build.


For the \nbLibVersionStr libraries of our dataset, \jdbl succeeds in producing a debloated \jar file for a total of \np{\nbLibDebSuccessNum} libraries, and fails to debloat \np{\nbDebloatError} libraries. Therefore, the overall debloating success rate of \jdbl  is \ShowPercentage{\nbLibDebSuccessNum}{\nbLibVersion}. 
When considering only the libraries that  originally compile, \jdbl succeeds in debloating \ShowPercentage{\nbLibDebSuccessNum}{\nbLibCompiling} of the libraries. 
We manually identify and classify the causes of failures in four categories:

\begin{itemize}[noitemsep]
\item \textit{Not compiled}. As a sanity-check, we compile the project before injecting \jdbl in its \mv build. The only modification consists in changing the \textit{pom.xml} to request the generation of a \jar that contains the bytecode of the project, along with all its runtime dependencies. If this step fails, the project does not compile, and it is ignored for the rest of the evaluation.
\item \textit{Crash.} We run a second \mv build, with \jdbl. This modifies the bytecode to remove unnecessary code. In certain situations, this procedure  causes the build to stop at some phase and terminate abruptly, \ie, due to accessing invalid memory addresses, using an illegal opcode, or triggering an unhandled exception.
\item \textit{Time-out.} \jdbl utilizes various coverage tools that instrument the bytecode of the project and its dependencies. This process induces an additional overhead in the \mv build process. Moreover, the incorrect instrumentation with at least one of the coverage tools may cause the test to enter into an infinite loop, \eg, due to blocking operations.
\item \textit{Validation error.} \mv includes dedicated plugins to check the integrity of the produced \jar file. \jdbl alters the behavior of the project build by packaging the debloated \jar using the \texttt{maven-assembly-plugin}. Some other plugins may not be compatible with \jdbl (\eg, when using customized assemblies), triggering validation errors during the build life-cycle. Moreover, we observe that for some libraries, the tests in the debloated \jar are not correctly executed due to particular library configurations in the \texttt{maven-surefire-plugin}.
\end{itemize}

We manually investigate the causes of the validation errors for the \nbLibValidation~libraries that fall into this category.
We found that \mv fails to validate the execution of the tests, either due to errors when running the instrumented code to collect coverage or incompatibilities among plugins that exercise the instrumented version of the library.
For example, in the case of \textit{org.apache.commons:\allowbreak collection:\allowbreak 4.0}, the \texttt{MANIFEST.MF} file is missing in the debloated \jar due to an incompatibility with library plugins. Therefore, \mv fails to package the debloated bytecode.
As another example, the \mv build of \textit{org.yaml:\allowbreak snakeyaml:\allowbreak 1.17} fails because of Yajta's instrumentation. 
This tool relies on Javassist for inserting probes in the bytecode. In this case, \jdbl changes a class that was frozen by Javassist when it was loaded. Consequently, Javassist crashes because further changes in a frozen class are prohibited.\footnote{\url{https://www.javassist.org/tutorial/tutorial.html}}\\

\begin{mdframed}[style=mpdframe]
\textbf{Answer to RQ1:} 
\jdbl successfully produces a debloated \jar file for \np{\nbLibDebSuccessNum} libraries in our dataset, which represents \ShowPercentage{\nbLibDebSuccessNum}{\nbLibCompiling} of the  libraries that compile correctly. 
This is the largest number of debloated subjects in the literature.
\end{mdframed}

\subsubsection{RQ2. \RQtwo}

Our second research question evaluates the behavior of the debloated library with respect to its original version. This evaluation is based on the test suite of the project. 
We investigate if the code debloated by \jdbl affects the results of the tests of the \np{\nbLibDebSuccessNum} libraries for which \jdbl produces a valid \jar file. 
This behavioral correctness assessment corresponds to the last phase in the execution of \jdbl.

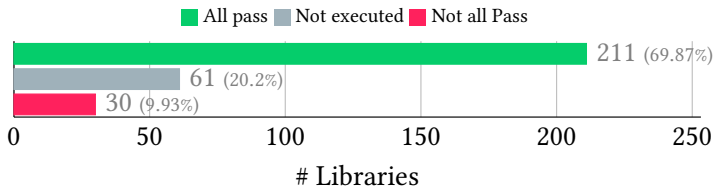
\begin{figure}
\centering
\resizebox{\columnwidth/2 + 3cm}{!}{\begin{tikzpicture} 
\begin{axis}[xbar,
ymin=0.5,
ymax=3.5,
xmin=0,
enlarge x limits={upper, value=0.2},
width=10.0cm,
height=2.5cm,
bar shift=0pt,
bar width=2.5mm,
axis x line* = bottom,
axis y line* = left,
xtick style={draw=none}, 
ytick style={draw=none}, 
yticklabels = \empty,
xmajorgrids = true,
xlabel=\# Libraries,
nodes near coords={\pgfmathprintnumber\myy~{\scriptsize(\pgfmathprintnumber\pgfplotspointmeta\%)}},
every node near coord/.append style={
    color = gray,
    rotate=0,
    anchor=west
},
visualization depends on=rawx \as \myy,
point meta={x*100/\nbLibDebSuccessNum},
legend style={
    row sep=3pt,
    draw=none,
    legend columns=-1,
    at={(0.5,1.6)},
    anchor=north,
    cells={anchor=west,font=\scriptsize}
},
legend image code/.code={
    \draw[#1, draw=none] (0cm,-0.1cm) rectangle (0.2cm,0.1cm);
},
]

\addplot+[pgreen] coordinates {(\nbLibPassTestNum,3)};
\addplot+[pgrey] coordinates {(\nbLibExcludedForTest,2)};
\addplot+[pred] coordinates {(\nbLibTestFailing,1)};

\addlegendentry{All pass}
\addlegendentry{Not executed}
\addlegendentry{Not all Pass}

\end{axis}
\end{tikzpicture}}
\caption{
	Number of debloated libraries for which the test suite passes; number of debloated libraries for which the number of executed tests does not match the original test execution (ignored for the research question); number of debloated libraries that have at least one failing test case.}
\label{fig:debloat_nb_test_errors_failures_stacked_barplot}
\end{figure}

\autoref{fig:debloat_nb_test_errors_failures_stacked_barplot} summarizes the comparison between the test suite executed on the original and the debloated libraries. 
From the \np{\nbLibDebSuccessNum} successfully debloated, \ChartSmall[p]{\nbLibPassTestNum}{\nbLibDebSuccessNum} preserve the original behavior (\ie, all the tests pass). In the case of \ChartSmall[p]{\nbLibTestFailing}{\nbLibDebSuccessNum} libraries, we observe at least one test failure.
This high test success rate is a fundamental result to ensure that the debloated version of the artifact preserves the behavior of the library. 
A table with the full list of the \nbLibPassTestNum ~successfully debloated libraries that pass all the tests is available in the replication package of this paper.\footnote{\url{https://github.com/castor-software/jdbl-experiments/blob/master/list_of_libs_succesfully_debloated_with_jdbl.md}} 

We excluded \ChartSmall[p]{\nbLibExcludedForTest}{\nbLibDebSuccessNum} libraries because the numbers of executed tests before and after the debloating did not match.
This is due to changes in the tests' configuration after injecting \jdbl into the build of the libraries. 
We excluded those libraries since different numbers of test runs imply a different test-based specification for the original and the debloated version of the library. Consequently, the results of the tests do not provide a sound basis for behavioral comparison.
The manual configuration of the libraries is a solution to handle this problem (expected usage of \jdbl), yet it is impractical in our experiments because of the large number of libraries that we debloat.

In total, we execute \np{\nbUniqueTestNum} unique tests, from which \np{\nbPassingTestNum} pass, and \np{\nbFailingTestNum} do not pass (\np{\nbFailureTest} fail, and \np{\nbErrorTest} result in error). This represents an overall behavior preservation ratio of~\np[\%]{\nbSuccessTestNum}, considering the total number of tests. This result shows that our code-coverage debloating approach is able to capture most of the project behavior, as observed by the tests, while removing the unnecessary bytecode.

We investigate the causes of test failures in the \np{\nbLibTestFailing} libraries that have at least one failure. To do so, we manually analyze the logs of the tests, as reported by \mv. We find the following \np{5} causes :

\begin{itemize}[noitemsep]
\item \texttt{NoClassDefFound (NCDF)}: \jdbl mistakenly removes a necessary class.
\item \texttt{TestAssertionFailure (TAF)}: the asserting conditions in the test fail for multiple reasons, \eg, flaky tests, or test configuration errors.
\item \texttt{UnsupportedOperationException (UOE)}: \jdbl mistakenly modifies the body of a necessary method, removing bytecode used by the test suite.
\item \texttt{NullPointerException (NPE)}: a necessary object is referenced before being instantiated.
\item \texttt{Other}: The tests are failing for another reason than the ones previously mentioned.
\end{itemize}

\autoref{tab:jdbl_failures_reasons} categorizes the tests failures for the \np{\nbLibTestFailing} libraries with at least one test that does not pass.
They are sorted in descending order according to the percentage of tests that fail on the debloated version.
The first column shows the name and version of the library.
Columns 2--7 represent the \np{5} causes of test failure according to our manual analysis of the tests' logs: \texttt{TAF}, \texttt{UOE}, \texttt{NPE}, \texttt{NCDF}, and \texttt{Other}.
The column labeled as \texttt{Other} shows the number of test failures that we were not able to classify.
The last column shows the percentage of tests that do not pass with respect to the total number of tests in each library. 
For example,  \textit{equalsverifier:3.4.1} has the largest number of test failures. After debloating, we observe \np{605} test failures out of \np{921} tests (\np{283} \texttt{TAF}, \np{221} \texttt{NCDF}, and \np{1} \texttt{Other}).
These test failures represent \ShowPercentage{605}{921} of the total number of tests in \textit{equalsverifier:3.4.1}.
This is an exceptional case, as for most of the debloated libraries, the tests that do not pass represent less than \np[\%]{5} of the total.


\begin{table}
\footnotesize
\centering
\caption{Classification of the tests that fail for the \np{\nbLibTestFailing} libraries that do not pass all the tests. We identified five causes of failures through the manual inspection of the Maven build testing logs:
TestAssertionFailure \texttt{(TAF)}, UnsupportedOperationException \texttt{(UOE)}, NullPointerException \texttt{(NPE)}, NoClassDefFound \texttt{(NCDF)}, and other unknown causes (\texttt{Other}).}
\label{tab:jdbl_failures_reasons}
\begin{tabularx}{0.8\textwidth}{@{}X|ccccc|r@{}}
\toprule
\multicolumn{1}{@{}l}{\textsc{Library}} &
\multicolumn{1}{l}{\begin{sideways}\texttt{TAF}\end{sideways}} & \multicolumn{1}{l}{\begin{sideways}\texttt{UOE}\end{sideways}} & 
\multicolumn{1}{l}{\begin{sideways}\texttt{NPE}\end{sideways}} & \multicolumn{1}{l}{\begin{sideways}\texttt{NCDF}\end{sideways}} &  
\multicolumn{1}{l}{\begin{sideways}\texttt{Other}\end{sideways}} &  
\multicolumn{1}{r}{\makecell{\textsc{Test failures}}} \\ 
\midrule
\textit{jai-imageio-core:1.3.1} &   &   &   & 3 &   & \ChartSmall[r]{3}{3} \\
\textit{jai-imageio-core:1.3.0} &   &   &   & 3 &   & \ChartSmall[r]{3}{3} \\
\textit{reflectasm:1.11.7} & 3 &   &   & 11 &   & \ChartSmall[r]{14}{16} \\
\textit{equalsverifier:3.3} & 273 &   &   & 315 & 1 & \ChartSmall[r]{589}{894} \\
\textit{equalsverifier:3.4.1} & 283 &   &   & 321 & 1 & \ChartSmall[r]{605}{921} \\
\textit{spark:2.0.0} & 3 &   &   &   & 22 & \ChartSmall[r]{25}{57} \\
\textit{logstash-logback-encoder:6.2} &   &   &   & 74 & 36 & \ChartSmall[r]{110}{307} \\
\bottomrule
\end{tabularx}
\end{table}

The most common cause of test failure is \texttt{NCDF} (\np{735}), followed by \texttt{TAF} (\np{592}). 
We found that these two types of failures are related to each other: when the test uses a non-covered class, the log shows a \texttt{NCDF}, and the test assertion fails consequently.
We notice that \texttt{NCDF} and \texttt{UOE} are directly related to the removal procedure during the debloating procedure, meaning that \jdbl is removing necessary classes and methods, respectively. 
This occurs because there are some Java constructs that \jdbl does not manage to cover dynamically, causing an incomplete debloating result, despite the union of information gathered from different coverage tools.
Primitive constants, custom exceptions, and single-instruction methods are typical examples. These are ubiquitous components of the Java language, which are meant to support robust object-oriented software design, with little or no procedural logic.
They are important for humans  and they are useless for the machine to run the program. Consequently, they are not part of the executable code in the bytecode, and cannot be covered dynamically.

\jdbl can generate a debloated program that breaks a few test cases. These cases reveal some limitations of \jdbl concerning behavior preservation, \ie, it fails to cover some classes and methods, removing necessary bytecode. 
One of the explanations is that the coverage tools modify the bytecode of the libraries. 
Those modifications can cause some test failures.
A failing test case stops the execution of the test and can introduce a truncated coverage report of the execution. Since some code is not executed after the failing assertion, some required classes or methods will not be covered and therefore debloated by \jdbl.
For example, in the \textit{reflections} library, a library that provides a simplified reflection API, some tests verify the number of fields of a class extracted by the library. 
However, JaCoCo injects a field in each class, which will invalidate the asserts of \textit{reflections} tests.

More generally, this reveals the remaining challenges of coverage-based debloating for real-world Java applications when using the test suite as workload. 
For this study, handling these challenging cases to achieve \np[\%]{100} correctness requires significant engineering effort, providing only marginal insights.
Therefore, we recommend always using our validation approach to be safe of semantic alterations when performing aggressive debloating transformations.\\

\begin{mdframed}[style=mpdframe]
\textbf{Answer to RQ2:} \jdbl automatically generates a debloated \jar that preserves the original behavior of \ChartSmall[p]{\nbLibPassTestNum}{\nbLibDebSuccessNum} libraries. 
A total of \np{\nbPassingTestNum} (\np[\%]{\nbSuccessTestNum}) tests pass on \np{\nbLibExecTestNum} libraries. 
This behavioral assessment of coverage-based debloating demonstrates that \jdbl preserves a large majority of the libraries' behavior, which is essential to meet the expectations of the libraries' users.
\end{mdframed}

\subsection{Debloating Effectiveness (RQ3, RQ4, and RQ5)}

In this section, we report on the effects of debloating Java libraries with \jdbl in terms of bytecode size reduction.

\subsubsection{RQ3. \RQthree}

To answer our third research question, we compare the status (kept or removed) of dependencies, classes, and methods in the \nbLibPassTest libraries correctly debloated with \jdbl. 
The goal is to evaluate the effectiveness of \jdbl to remove these bytecode elements through coverage-based debloating. 

\begin{figure*}
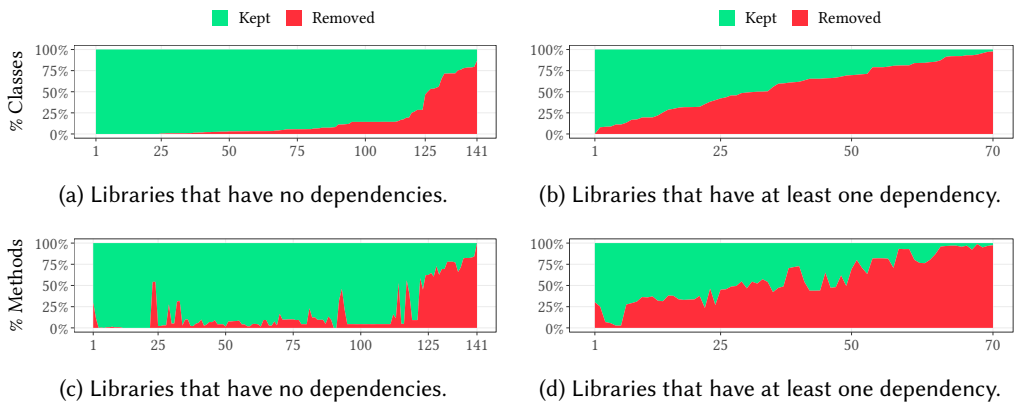

\centering
\subcaptionbox{Libraries that have no dependencies.\label{fig:3a}}{%
	\resizebox{\columnwidth/2-0.25cm}{!}{\input{lib_classes_area.tex}}
}
\vspace{0.25cm}
\subcaptionbox{Libraries that have at least one dependency.\label{fig:3b}}{%
	\resizebox{\columnwidth/2-0.25cm}{!}{\input{deps_classes_area.tex}}
}\\
\subcaptionbox{Libraries that have no dependencies.\label{fig:3c}}{%
	\resizebox{\columnwidth/2-0.25cm}{!}{\input{lib_methods_area.tex}}
}
\subcaptionbox{Libraries that have at least one dependency.\label{fig:3d}}{%
	\resizebox{\columnwidth/2-0.25cm}{!}{\input{deps_methods_area.tex}}
}
\caption{Percentage of classes kept and removed in (a) libraries that have no dependencies, and (b) libraries that have at least one dependency. Percentage of methods kept and removed in (c) libraries that have no dependencies, and (d) libraries that have at least one dependency.}
\label{fig:debloat_lib_classes_methods_areas}
\end{figure*}

\autoref{fig:debloat_lib_classes_methods_areas} shows area charts representing the distribution of kept and removed classes and methods in the \nbLibPassTest correctly debloated libraries.
To analyze the impact of dependencies, we separate the libraries into two sets: the libraries that have no dependency (Figures~\ref{fig:3a} and~\ref{fig:3c}), and the libraries that have at least one dependency (Figures~\ref{fig:3b} and~\ref{fig:3d}).
In each figure, the x-axis represents the libraries in the set, sorted in increasing order according to the number of removed classes, whereas the y-axis represents the percentage of classes (Figures~\ref{fig:3a} and~\ref{fig:3b}) or methods (Figures~\ref{fig:3c} and~\ref{fig:3d}) kept and removed.
The order of the libraries,  on the x-axis, is the same for each figure.

\autoref{fig:3a} shows the comparison between the percentages of kept and removed classes in the \np{\nbLibWithoutDependencies} libraries that have no dependency. 
A total of \np{116} libraries have at least one removed class. 
The library with the largest percentage of removed classes is \textit{jfree-jcommon:1.0.23} with the \np[\%]{86.7} of its classes considered as bloated. 
On the other hand, \autoref{fig:3b} shows the percentage of removed classes for the \np{\nbLibWithDependencies} libraries that have at least one dependency. 
We observe that the ratio of removed classes in these libraries is significantly higher with respect to the libraries with no dependencies.
All the libraries that have dependencies have at least one removed class, and \np{45} libraries have more than \np[\%]{50} of their classes bloated. 
This result hints at the importance of reducing the number of dependencies to mitigate software bloat. 

\autoref{fig:3c} shows the percentage of kept and removed methods in the \np{\nbLibWithoutDependencies} libraries that have no dependencies. 
We observe that libraries with a few removed classes still contain a significant percentage of removed methods.
For example, the library \textit{net.iharder:base64:2.3.9} has \np[\%]{42.2} of its methods removed in the \np[\%]{99.4} of its kept classes. 
This suggests that a fine-grained debloat, to the level of methods, is beneficial for some libraries. The used classes may still contain a significant number of bloated methods. 
On the other hand, Figure~\ref{fig:3d} shows the percentage of kept methods in libraries with at least one dependency. 
All the libraries have a significant percentage of removed methods. 
As more bloated classes are in the dependencies, the artifact globally includes more bloated methods. 

\begin{figure}[t]
\centering
\includegraphics[width=0.6\columnwidth]{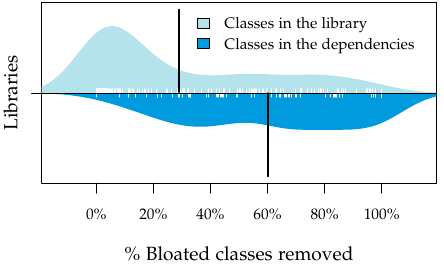}
\caption{Distribution of the percentage of bloated classes that belong to libraries, and bloated classes that belong to dependencies. The strip chart (white marks) in between represents the libraries that belong to each of the two groups. The two vertical bars represent the average value for each group.}
\label{fig:bloat_libs_deps_bean}
\end{figure}

Now we focus on determining the difference between the bloat that is caused exclusively by the classes in the library, and the bloat that is a consequence of software reuse through the declaration of dependencies.
\autoref{fig:bloat_libs_deps_bean} shows a beanplot~\cite{Kampstra2008} comparing the distribution of the percentage of bloated classes in libraries, with respect to the bloated classes in dependencies. 
The density shape at the top of the plot shows the distribution of the percentage of bloated classes that belong to the \nbLibPassTest libraries. 
The density shape at the bottom shows this percentage for the classes in the dependencies of the \np{\nbLibWithDependencies} libraries that have at least one dependency.
The average bloat in libraries is \np[\%]{27.3}, whereas in the dependencies it is \np[\%]{59.8}. 
Overall, the average of bloated classes removed for all the libraries, including their dependencies, is \np[\%]{32.5}.
We perform a two-samples Wilcoxon test, which confirms that there are significant differences between the percentage of bloated classes in the two groups (p-value $ < 0.01$).
Therefore, we reject the null hypothesis and confirm that 
the ratio of bloated classes is more significant among the dependencies than among the classes of the artifacts.\looseness=-1

\begin{table}[t]
\footnotesize
\centering
\caption{Summary of the number of dependencies, classes, and methods removed in the \nbLibPassTest ~libraries correctly debloated with \jdbl.}
\label{tab:bloat_summary}
\begin{tabular}{@{}lr@{}}
\toprule
             & \textsc{Removed (\%)} \\ \midrule

Dependencies & \ChartSmall[r]{\nbBloatedDependencies}{\nbDependencies}     \\
Classes      & \ChartSmall[r]{\nbBloatedClasses}{\nbClasses}  \\
Methods      & \ChartSmall[r]{\nbBloatedMethods}{\nbMethods} \\
\bottomrule
\end{tabular}
\vspace{-0.2cm}
\end{table}

\autoref{tab:bloat_summary} summarizes the debloating results for the dependencies, classes, and methods. Interestingly, \jdbl completely removes the bytecode for \ShowPercentage{\nbBloatedDependencies}{\nbDependencies} of the dependencies.
In other words, \nbBloatedDependencies\xspace  dependencies in the dependency tree of the projects are not necessary to successfully execute the workload in our dataset.
At the class level, we find \ShowPercentage{\nbBloatedClasses}{\nbClasses} of bloat, from which we determine that \ShowPercentage{\nbBloatedDepClasses}{\nbBloatedClasses} belong to dependencies.
\jdbl debloats \ShowPercentage{\nbBloatedMethods}{\nbMethods} of the methods, from which  \ShowPercentage{\nbBloatedDepMethods}{\nbBloatedMethods} belong to dependencies.\\

\begin{mdframed}[style=mpdframe]
\textbf{Answer to RQ3:} \jdbl removes bytecode in all libraries. It reduces the number of dependencies, classes, and methods by \ShowPercentage{\nbBloatedDependencies}{\nbDependencies}, \ShowPercentage{\nbBloatedClasses}{\nbClasses}, and \ShowPercentage{\nbBloatedMethods}{\nbMethods}, respectively. This result confirms the relevance of the coverage-based debloating approach for reducing the unnecessary bytecode of Java projects, while preserving their correctness.
\end{mdframed}

\subsubsection{RQ4. \RQfour}
\label{sec:rq4}

We consider all the elements in the \jar files before the debloating, and study the size of the debloated version of the artifact, with respect to the original bundle.
Decreasing the size of \jar files by removing bloated bytecode has a positive impact on saving space on disk, and helps reduce overhead when the \jar files are shipped over the network.

\jar files contain bytecode, as well as additional resources that depend on the functionalities of the artifacts (\eg, \texttt{HTML}, \texttt{DLL}, \texttt{SO}, and \texttt{CSS} files). 
\jar files also contain resources required by \mv to handle configurations and dependencies (\eg, \texttt{MANIFEST.MF} and \textit{pom.xml}).
However, \jdbl is designed to debloat only executable code (\texttt{class} files). Therefore, we assess the impact of bytecode removal with respect to the executable code in the original bundle.

\autoref{tab:size_impact} summarizes the main metrics related to the content and size of the \jar files in our dataset. 
We observe that the additional resources represent \ShowPercentage{\resourceSize}{\totalSize} of the total \jar size, whereas \ShowPercentage{\bytecodeSize}{\totalSize} of the size is dedicated to the bytecode.
This observation supports the relevance of debloating the bytecode in order to shrink the size of the Maven artifacts. 

\begin{table}[t]
\footnotesize
    \centering
    \begin{tabularx}{.45\textwidth}{@{}X r@{}}
    \toprule
\textsc{Metrics}        & \textsc{Size in bytes (\%)}             \\ 
\midrule
Resources  & \ChartSmall{\resourceSize}{\totalSize}         \\
Bytecode   & \ChartSmall{\bytecodeSize}{\totalSize}         \\ 
\midrule
Non-bloated classes   & \ChartSmall{\nonBloatedSize}{\bytecodeSize}    \\
Bloated classes        & \ChartSmall{\debloatedSize}{\bytecodeSize}     \\
Bloated methods & \ChartSmall{\debloatedMethodSize}{\bytecodeSize}\\
\midrule
Total size     & \np{\totalSize} \\
\bottomrule
    \end{tabularx}
    \caption{Size in bytes of the elements in the JAR files.}
    \label{tab:size_impact}
\end{table}

Overall, the bloated elements in the compiled artifacts in our dataset represent {\np{610.3}/\np[MB]{893.7}} (\ShowPercentage{\bloatedSize}{\bytecodeSize}) of pure bytecode: {\np[MB]{596.5}} of bloated classes and {\np[MB]{13.8}} of bloated methods. 
The used bytecode represents  \ShowPercentage{\nonBloatedSize}{\bytecodeSize} of the size.
Interfaces, enumeration types, annotations, and exceptions represent $15.8\%$ of the size on disk of all the \texttt{class} files.
In comparison with the classes, the debloat of methods represents a relatively limited size reduction. 
This is because we are reporting the removal of methods in the classes that are not entirely removed by \jdbl.
Furthermore, the methods cannot be completely removed, only the body of the method is replaced by an exception as detailed in \autoref{sec:approach}.

\begin{figure}[t]
\centering
\includegraphics[width=0.6\columnwidth]{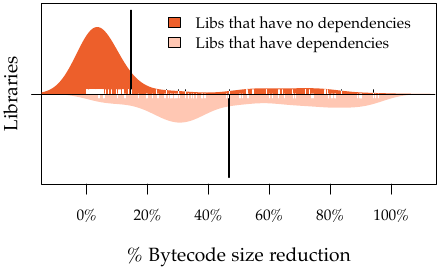}
\caption{Distribution of the percentage of reduction of the \jar size in libraries that have no dependencies and libraries that have at least one dependency, with respect to the original bundle.}
\label{fig:lib_size_violinplot}
\end{figure}

\autoref{fig:lib_size_violinplot} shows a beanplot comparing the distribution of the percentage of bytecode reduction in the libraries that have no dependency, with respect to the libraries that have at least one dependency.
From our observations, the average bytecode size reduction in the libraries that have dependencies (\np[\%]{46.7}) is higher than the libraries with no dependencies (\np[\%]{14.5}).
Overall, the average percentage of bytecode reduction for all the libraries is \np[\%]{25.8}.
We performed a two-samples Wilcoxon test, which shows that there are significant differences between those two groups (p-value $ < 0.01$). 
Therefore, we reject the null hypothesis that the coverage-based debloating approach has the same impact in terms of reduction of the \jar size for libraries that declare dependencies, and libraries that do not.\looseness=-1

We perform a Spearman's rank correlation test between the original number of classes in the libraries and the size of the removed bytecode. We found that there is a significant positive correlation between both variables ($\rho$ =$0.97$, p-value $< 0.01$).
This result confirms the intuition that projects with many classes tend to occupy more space on disk due to bloat. 
However, the decision of what is necessary or not heavily depends on the library, as well as on the workload.\\

\begin{mdframed}[style=mpdframe]
\textbf{Answer to RQ4:} \jdbl removes \np[\%]{68.3} of pure bytecode in \jar files, which represents an average size reduction of   \np[\%]{25.8} per library \jar file.
The \jar size reduction is significantly higher in libraries with at least one dependency compared to libraries with no dependency. 
\end{mdframed}

\subsubsection{RQ5. \RQfive}
\label{sec:rq5}

In this research question, we compare the debloating results of \jdbl with respect to \jshrink. The comparison is based on two metrics: the reduction of bytecode size after debloat, and the preservation of test results after debloat.
The \jshrink benchmark includes \np{25} Java projects. To answer RQ5, we discard \np{8} projects: \np{6} that are multi-module (\jdbl is designed to debloat single-module Maven projects), and \np{2} projects whose builds fail due to test errors caused by unavailable network resources.
Therefore, our comparison of \jdbl and \jshrink is based on $17$  projects in total.
For each project, we configure it to execute \jdbl and generate a debloated version of the fat \jar.
To validate the semantic correctness of the debloated artifact, we execute the project's test suite on the debloated version.

\autoref{tab:jshrink_benchmark} describes the benchmark along with the debloating results obtained with \jshrink and \jdbl. 
The first column shows the name of the project as it appears on GitHub.
The second column shows the commit SHA of the project, which is the same used in the companion paper of \jshrink~\cite{Bruce2020}.
The third column is the size of the original packaged \jar of the project, which includes all its dependencies with a \texttt{compile} scope.
Projects are sorted in decreasing order according to their original size. 

We report the bytecode size reduction of the debloated version of the projects achieved with \jshrink and \jdbl.
The average size reduction achieved with \jdbl is \np[\%]{35.1}, which is more than double the size reduction obtained with \jshrink. 
An explanation is that \jshrink makes a more conservative debloating decision by setting all public methods, main methods, and test methods of libraries as entry points to approximate possible usages, whereas \jdbl debloats according to a workload (\ie, the coverage information collected by running the test suite of the library).
We observe that the percentage of reduction varies greatly among the projects depending on their size.
We performed a Spearman’s rank correlation test between the size of the original compiled project and the percentage of size reduction obtained with \jshrink and \jdbl.
We found that there is a significant positive correlation between both variables for \jshrink ($\rho=0.52$, p-value < $0.05$) and \jdbl ($\rho=0.52$, p-value < $0.05$). 
This result confirms the results obtained in RQ4 where we show that larger libraries are prone to become more bloated.

\input{jshrink_benchmark}

To assess the behavior preservation of the debloated projects \wrt their original version, we run existing test cases before and after debloating.
\autoref{tab:jshrink_benchmark} shows the semantic preservation capabilities of \jshrink and \jdbl on the benchmark of \np{17} projects.
We consider a debloated project to have broken semantics if at least one of its tests fails after debloating. 
A project with no broken semantic is denoted by \textcolor{teal}{\ding{51}}, while \textcolor{red}{\ding{55}} denotes the presence of at least one test failure after debloating.
\jshrink causes test failures in \np{3} projects (\np{77} failures in total).
On the contrary, \jdbl preserves the behavior of all the projects on this benchmark, according to the results of the tests.\\

\begin{mdframed}[style=mpdframe]
\textbf{Answer to RQ5:} \jdbl successfully debloats the \np{17} single-module Java projects in the benchmark of Bruce \etal~\cite{Bruce2020}, with a size reduction of \np[\%]{35.1} on average, and preserves the behavior according to the tests. JShrink reduces size by \np[\%]{15.1} on average. This is evidence that coverage-based debloating is a promising technique that advances the state-of-the-art of Java bytecode debloating. 
\end{mdframed}

\subsection{Impact of Debloating on Library Clients (RQ6 and RQ7)}

In this section, we study the repercussion of performing coverage-based debloating on library clients.
To the best of our knowledge, this is the first experimental report that  measures the impact of debloating libraries on the syntactic and semantic correctness of their clients.

\subsubsection{RQ6. \RQsix}

\begin{table*}[t]
    \centering
\footnotesize
    \begin{tabularx}{1.0\textwidth}{@{}Xrrr@{}}
     \toprule
\textsc{Description} & \textsc{\# Libraries} & \textsc{\# Clients} & \textsc{Occurrence} \\\midrule

Cannot find class & \ChartSmall[q]{12}{19} & \ChartSmall[q]{20}{38} & \ChartSmall[q]{314}{640} \\
Unmappable character for encoding \texttt{UTF8} & \ChartSmall[q]{1}{19} & \ChartSmall[q]{1}{38} & \ChartSmall[q]{100}{640} \\
Package does not exist & \ChartSmall[q]{6}{19} & \ChartSmall[q]{13}{38} & \ChartSmall[q]{78}{640} \\
Cannot find variable & \ChartSmall[q]{7}{19} & \ChartSmall[q]{9}{38} & \ChartSmall[q]{77}{640} \\
Cannot find method & \ChartSmall[q]{1}{19} & \ChartSmall[q]{1}{38} & \ChartSmall[q]{28}{640} \\
Static import only from classes and interfaces & \ChartSmall[q]{1}{19} & \ChartSmall[q]{1}{38} & \ChartSmall[q]{14}{640} \\
Method does not override a method from a supertype & \ChartSmall[q]{2}{19} & \ChartSmall[q]{2}{38} & \ChartSmall[q]{12}{640} \\
Processor error & \ChartSmall[q]{2}{19} & \ChartSmall[q]{11}{38} & \ChartSmall[q]{11}{640} \\
Cannot find other symbol & \ChartSmall[q]{2}{19} & \ChartSmall[q]{2}{38} & \ChartSmall[q]{4}{640} \\
UnsupportedOperationException & \ChartSmall[q]{1}{19} & \ChartSmall[q]{1}{38} & \ChartSmall[q]{1}{640} \\
Plugin verification & \ChartSmall[q]{1}{19} & \ChartSmall[q]{1}{38} & \ChartSmall[q]{1}{640} \\
\midrule
\np{11} Unique errors & \np{19} Unique libraries & \np{38} Unique clients & \np{640} Compilation errors \\
\bottomrule
    \end{tabularx}
    \caption{Frequency of the errors for the \np{\nbClientDebloatError} unique clients that have compilation errors. Note that a client can have multiple errors from different categories.}
    \label{tab:compilation_error}
\end{table*}

In this research question, we investigate how debloating a library with a coverage-based approach impacts the compilation of the library's clients. 
We hypothesize that the essential functionalities of the library are less likely to be debloated, limiting the syntactic negative impact on their clients. 

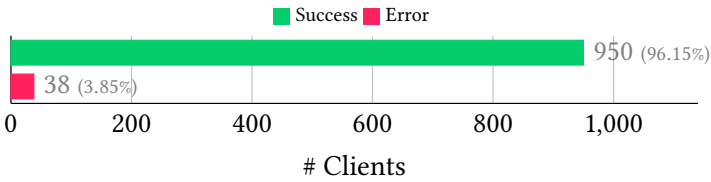
\begin{figure}
\centering
\resizebox{\columnwidth/2 + 3cm}{!}{\begin{tikzpicture} 
\begin{axis}[xbar,
ymin=0.5,
ymax=2.5,
xmin=0,
enlarge x limits={upper, value=0.2},
width=10.0cm,
height=2.4cm,
bar shift=0pt,
bar width=3mm,
axis x line* = bottom,
axis y line* = left,
xtick style={draw=none}, 
ytick style={draw=none}, 
yticklabels = \empty,
xmajorgrids = true,
xlabel=\# Clients,
nodes near coords={\pgfmathprintnumber\myy~{\scriptsize(\pgfmathprintnumber\pgfplotspointmeta\%)}},
every node near coord/.append style={
    color = gray,
    rotate=0,
    anchor=west
},
visualization depends on=rawx \as \myy,
point meta={x*100/\nbStatCoveringClient},
legend style={
    row sep=3pt,
    draw=none,
    legend columns=-1,
    at={(0.5,1.6)},
    anchor=north,
    cells={anchor=west,font=\scriptsize}
},
legend image code/.code={
    \draw[#1, draw=none] (0cm,-0.1cm) rectangle (0.2cm,0.1cm);
},
]

\addplot+[pgreen] coordinates {(\FPprint{\nbStatCoveringClient-\nbClientDebloatError},2)};
\addplot+[pred] coordinates {(\nbClientDebloatError,1)};

\addlegendentry{Success}
\addlegendentry{Error}

\end{axis}
\end{tikzpicture}}
\caption{
	Results of the compilation of the \np{\nbStatCoveringClient} clients that use at least one debloated library in the source code.}
\label{fig:client_compilation_result}
\end{figure}

As described in \autoref{sec:rq5_protocol},  we consider the \np{\nbClient} clients that use the \np{\nbLibPassTestNum} debloated libraries that pass all the tests. 
We check that the clients use at least one class in the library through static analysis.
We identify \ChartSmall[q]{\nbStatCoveringClient}{\nbClient} clients that satisfy this condition.
\autoref{fig:client_compilation_result} shows the results obtained after attempting to compile the clients with the debloated library.
\jdbl generates debloated  libraries for which \ChartSmall[p]{\nbCientDebloatNoCompError}{\nbStatCoveringClient} of their clients successfully compile.

From the \np{\nbStatCoveringClient} clients that use at least one class of the library, we only observe compilation failures in \ChartSmall[p]{\nbClientDebloatError}{\nbStatCoveringClient} clients.
\autoref{tab:compilation_error} shows our manual classification of the errors, based on the analysis of the \mv build logs. 
The first column describes the error message, columns 2--3 represent the number of libraries that trigger this kind of error, and the number of clients that are affected and the percentage relative to the number of libraries or clients impacted by a compilation error. 
Note that a client can be impacted by several different errors. 
The fourth column represents the occurrence of the error in the clients, as quantified from the \mv logs.

The causes of compilation errors are diverse.
However, they are primarily due to errors related to missing packages, classes, methods, and variables (\ShowPercentage{511}{640} of all the compilations errors).
The debloating procedure directly causes those errors, as  the bytecode elements are removed, and  the clients do not compile. 
We detect \np{640} errors in total. Most of them occur for similar causes.
Indeed, \ChartSmall[q]{20}{38} clients are not compiling because of one single error cause (which can be unique for each client).
Moreover, when a client fails for a library, the other clients of the same library are generally failing for the same reason. 
It means that a single action can solve most of the client's problems, \ie, by adding the missing element to the debloated library.

Several clients face compilation issues because of their plugins. In order to have the client  use the debloated library, we inject the debloated library inside the bytecode folder of the clients. 
Unfortunately, some plugins of the clients will also analyze the bytecode of the debloated library that may not follow the same requirements.
\texttt{Plugin verification error}, \texttt{Unmappable character for encoding UTF8}, and \texttt{Processor error} are related to this type of bytecode validation.

In the list of errors, we also observe a runtime exception: \texttt{UnsupportedOperationException}.
This error is unexpected since the compilation should not execute code and therefore should not trigger a runtime exception. 
It happens during the build of \href{https://github.com/jenkinsci/warnings-plugin}{jenkinsci/warnings-plugin}, which uses the  \textit{commons-io:2.6} library. 
In this case, the compilation itself of this client does not fail, but the  \mv build does. 
One of the \mv plugins of this project relies on one method that is debloated in \texttt{apache/commons-io}, therefore, the compilation does not fail because of the source code of the client but because of one particular \mv plugin used by the client.\\

\begin{mdframed}[style=mpdframe]
\textbf{Answer to RQ6:} \jdbl preserves the syntactic correctness of \ChartSmall[p]{\nbCientDebloatNoCompError}{\nbStatCoveringClient} clients that use a library debloated by \jdbl. 
This is the first empirical demonstration that debloating can preserve essential functionalities to successfully compile the clients of debloated libraries.
\end{mdframed}

\subsubsection{RQ7. \RQseven}

In this research question, we analyze another facet of debloating that may affect the clients: the disturbance of their behavior.
To do so, we use the test suite of the clients as the oracle for assessing correct behavior.
If a test in the client fails with the debloated library, then the coverage-based debloating breaks the behavior of the client.

We consider \np{\nbClient} clients to answer this question. They  use the \np{\nbLibPassTestNum} debloated libraries that pass all the tests. 
We check that the client tests cover at least one class in the library, through dynamic code coverage.
We identify \ChartSmall[q]{\nbDynCoveringClient}{\nbClient} clients that satisfy this condition.

\begin{figure}[t]
\centering
\resizebox{\columnwidth/2 + 3cm}{!}{\begin{tikzpicture} 
\begin{axis}[xbar,
ymin=0.5,
ymax=2.5,
xmin=0,
enlarge x limits={upper, value=0.2},
width=10.0cm,
height=2.4cm,
bar shift=0pt,
bar width=3mm,
axis x line* = bottom,
axis y line* = left,
xtick style={draw=none}, 
ytick style={draw=none}, 
yticklabels = \empty,
xmajorgrids = true,
xlabel=\# Clients,
nodes near coords={\pgfmathprintnumber\myy~{\scriptsize(\pgfmathprintnumber\pgfplotspointmeta\%)}},
every node near coord/.append style={
    color = gray,
    rotate=0,
    anchor=west
},
visualization depends on=rawx \as \myy,
point meta={x*100/\nbDynCoveringClient},
legend style={
    row sep=3pt,
    draw=none,
    legend columns=-1,
    at={(0.5,1.6)},
    anchor=north,
    cells={anchor=west,font=\scriptsize}
},
legend image code/.code={
    \draw[#1, draw=none] (0cm,-0.1cm) rectangle (0.2cm,0.1cm);
},
]

\addplot+[pgreen] coordinates {(\FPprint{\nbDynCoveringClient-\nbFailingTestClient},2)};
\addplot+[pred] coordinates {(\nbFailingTestClient,1)};

\addlegendentry{All tests pass}
\addlegendentry{Not all tests pass}

\end{axis}
\end{tikzpicture}}
\caption{
	Results of the tests on the \np{\nbDynCoveringClient} clients that cover at least one debloated library.}
\label{fig:client_test_result}
\end{figure}
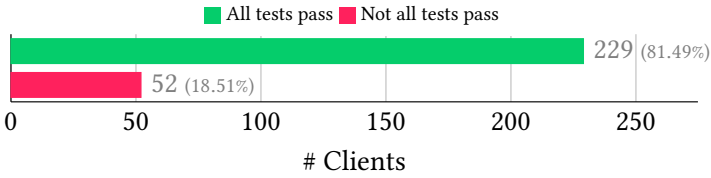

\autoref{fig:client_test_result} presents the test results after building the clients with the debloated library.
In total, \ChartSmall[p]{\nbClientDebloatNoTestFail}{\nbDynCoveringClient} clients pass all their tests, \ie, they behave the same with the original and with the debloated library. 
There are \ChartSmall[p]{\nbFailingTestClient}{\nbDynCoveringClient} clients that have more failing test cases with the debloated library than with the original.
However, the number of tests that fail is only \ChartSmall[q]{\nbClientFailure}{\nbClientTest} of the total number of tests in the clients.
This  indicates that the negative impact of debloated libraries, as measured by the number of affected client tests, is marginal.

\begin{table*}[t]
    \centering
\footnotesize
    \begin{tabularx}{0.85\textwidth}{@{}Xrrr@{}}
    \toprule
\textsc{Exception} & \textsc{\# Libraries} & \textsc{\# Clients} & \textsc{Occurrence} \\\midrule

UnsupportedOperationException & \ChartSmall[q]{27}{\nbFailingTestProject} & \ChartSmall[q]{40}{\nbFailingTestClient} & \ChartSmall[q]{609}{\nbClientFailure} \\
IllegalStateException & \ChartSmall[q]{1}{\nbFailingTestProject} & \ChartSmall[q]{2}{\nbFailingTestClient} & \ChartSmall[q]{55}{\nbClientFailure} \\
NoClassDefFoundError & \ChartSmall[q]{6}{\nbFailingTestProject} & \ChartSmall[q]{6}{\nbFailingTestClient} & \ChartSmall[q]{24}{\nbClientFailure} \\
Assert & \ChartSmall[q]{5}{\nbFailingTestProject} & \ChartSmall[q]{5}{\nbFailingTestClient} & \ChartSmall[q]{12}{\nbClientFailure} \\
ExceptionInInitializerError & \ChartSmall[q]{4}{\nbFailingTestProject} & \ChartSmall[q]{4}{\nbFailingTestClient} & \ChartSmall[q]{4}{\nbClientFailure} \\
TargetHostsLoadException & \ChartSmall[q]{1}{\nbFailingTestProject} & \ChartSmall[q]{1}{\nbFailingTestClient} & \ChartSmall[q]{4}{\nbClientFailure} \\
NullPointerException & \ChartSmall[q]{1}{\nbFailingTestProject} & \ChartSmall[q]{1}{\nbFailingTestClient} & \ChartSmall[q]{3}{\nbClientFailure} \\
TimeoutException & \ChartSmall[q]{1}{\nbFailingTestProject} & \ChartSmall[q]{1}{\nbFailingTestClient} & \ChartSmall[q]{2}{\nbClientFailure} \\
PushSenderException & \ChartSmall[q]{1}{\nbFailingTestProject} & \ChartSmall[q]{1}{\nbFailingTestClient} & \ChartSmall[q]{2}{\nbClientFailure} \\
AssertionError & \ChartSmall[q]{1}{\nbFailingTestProject} & \ChartSmall[q]{1}{\nbFailingTestClient} & \ChartSmall[q]{1}{\nbClientFailure} \\
\midrule
\np{10} Unique exceptions & \np{\nbFailingTestProject} Unique libraries & \np{\nbFailingTestClient} Unique clients & \np{\nbClientFailure} Failing tests \\

\bottomrule
    \end{tabularx}
    \caption{Frequency of the exceptions thrown during the execution of the tests for the \np{\nbFailingTestClient} unique clients that have failing test cases. Note that a client can have multiple exceptions from different categories.}
    \label{tab:client_test_failures}
\end{table*}

We investigate the causes of the test failures.
\autoref{tab:client_test_failures} quantifies the exceptions thrown by the clients.
The first column shows the \np{10} types of exceptions that we find in the Maven logs. Columns 2--3 represent the number of libraries involved in the failure and the number of clients affected. 
Column 4 represents the occurrence of the exception, as quantified from the logs.

\texttt{UsupportedOperationException} is the most frequent exception, with \np{609} occurrences in the failing-tests,  affecting \ShowPercentage{40}{\nbFailingTestClient} of clients with errors.
This exception is triggered when one of the debloated methods is executed. 
The second most common exception is \texttt{Illegal\allowbreak{}State\allowbreak{}Exception}, with a total of \np{55} occurrences. 
This exception happens when the client tries to load a bloated configuration class and fails.
The third most frequent exception, with \np{24} occurrences, is \texttt{No\allowbreak{}Class\allowbreak{}Def\allowbreak{}FoundError}.
Similar to \texttt{Usupported\allowbreak{}Operation\allowbreak{}Exception}, it happens when the clients try to load a debloated class in the library.

Interestingly, there are only \np{12} assertion related exceptions (\np{11} \texttt{Assert} and \np{1} \texttt{AssertionError}). 
Having runtime exceptions that are triggered during the executions of the clients is less harmful than having behavior changes. 
The runtime exceptions can be monitored and handled while running the client, while a behavior change can stay hidden and corrupt the state of the clients.\\ 

\begin{mdframed}[style=mpdframe]
\textbf{Answer to RQ7:} \jdbl preserves the behavior of \ChartSmall[p]{\nbClientDebloatNoTestFail}{\nbDynCoveringClient} clients of debloated libraries.
The \np{\nbFailingTestClient} other clients still pass \ChartSmall[p]{43684}{\nbClientTest} of their test cases. In these cases, \np[\%]{99.1} of the test failures are due to a missing class or method, which can be easily located and fixed.
This experiment shows that the risks of removing code in software libraries are limited for their clients.\looseness=-1
\end{mdframed}

\section{Discussion}\label{sec:threats}

In this section, we discuss key aspects of the design for coverage-based debloating. Then, we address the threats to the validity of the evaluation of \jdbl.

\subsection{Complementarity of Code-Coverage Tools}

As presented in \autoref{sec:approach}, we leverage the diversity of implementations of code coverage tools and the dynamic logging capabilities of the JVM class loader to collect precise coverage information.  
Now, we discuss the advantages of using this approach for debloating.

\begin{figure}[t]
\centering
\includegraphics[width=5.5cm]{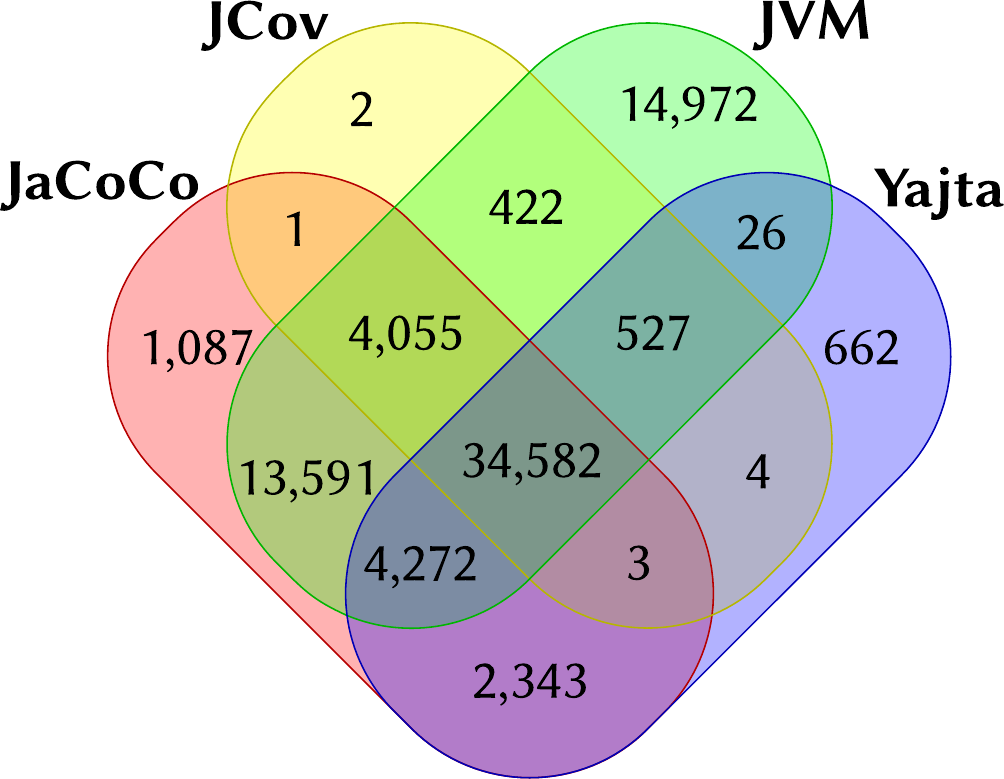}
\caption{Number of classes covered by the diverse tools used to collect coverage information in our dataset of \nbLibPassTestNum~successfully debloated libraries.}
\vspace{-0.2cm}
\label{fig:venn_coverage_tools}
\end{figure}

We collect and aggregate the coverage reports of the four tools used by \jdbl to capture class usage information: JaCoCo, JCov, Yajta, and the JVM class loader.
We consider a class as covered if it is reported as used by at least one of these tools.
\autoref{fig:venn_coverage_tools} shows a Venn diagram of the classes reported as covered by each tool.
There is a total of \np{76549} classes covered by at least one tool in our dataset of \nbLibPassTestNum ~successfully debloated libraries. 
One key observation is that JaCoCo covers only \ChartSmall[p]{59934}{76549} of the classes that are used when running our workloads. This means that if we rely on JaCoCo only, the state-of-the-art coverage tool for Java, we would capture a significant share of false positive cases of bloated classes. This is critical, since removing these classes would produce a debloated project that cannot be properly used to run the workload.
Another interesting observation is about the diversity of behaviors in modern coverage tools. There are only \ChartSmall[p]{34582}{76549} classes that are covered by the four tools.
The JVM class loader is the one that captures the largest number of unique classes: \ChartSmall[p]{14972}{76549}. 
This is because it logs the usage of dynamically loaded classes at runtime. 
In contrast, the other coverage tools can provide more fine-grained coverage information (\eg, methods and instruction) but miss usages of dynamically loaded resources  (\eg, classes loaded via the Java reflection mechanism). 
The addition of JCov and Yajta improves the coverage of used classes, accounting for $2$ and $662$ unique covered classes, respectively.


This is evidence that combining the features of several coverage tools is relevant to accurately capture the code that is used at runtime.
Capturing the complete coverage of classes that are necessary for a workload is critical for debloating.
Failing to do so leads to the generation of a debloated project that does not compile, or even worse, that  leads to runtime errors when client projects use debloated libraries. 
A more in-depth analysis of the similarities and differences of code coverage tools is a promising direction for future research on Java bytecode debloating.

\subsection{Execution Time}

The vision for debloating is to integrate it as part of building pipelines. In this context, execution time is an important consideration. Here, we discuss whether the execution time of \jdbl makes  coverage-based debloating  feasible for everyday software development.
In our experiments, we executed \jdbl on $395$ libraries, for a total of 1 day and 10:55:00h. 
This represents an average debloating time of $5.3$ minutes per library.
\autoref{tab:benchmark_execution_time} shows execution times of \jdbl and \jshrink, with the benchmark of Bruce \etal~\cite{Bruce2020}.
The first and second columns show the projects' names and commit SHAs, respectively. 
The third and fourth columns show the execution time of \jdbl and \jshrink for each project, measured in seconds and sorted in decreasing order according to the results of \jdbl. 
The comparison is made on the same hardware as the main \jdbl experiment, described in \autoref{sec:protocol}.
We observe that it took a total of \np{2685} seconds (less than one hour) to debloat the benchmark with \jdbl, whereas \jshrink took a total of \np{29397} seconds (more than eight hours). 
The average debloating time for this benchmark using \jdbl is $157.9s$ (less than $3$ minutes per project), which is $11$ times faster than \jshrink.

With times in the range of minutes,  coverage-based debloating with \jdbl can be used in a daily build. We also show that \jdbl significantly improves the state-of-the-art of Java debloating, regarding execution time. This is an important contribution toward the integration of debloating into regular development processes.  

\begin{table*}[t]
\scriptsize
\centering
\caption{Execution time of \jdbl and \jshrink in the benchmark of Bruce \etal~\cite{Bruce2020}.\\}
\label{tab:benchmark_execution_time}
\begin{tabular}{lccc}

\toprule
&  & \textsc{\jdbl} & \textsc{\jshrink}\\
\textsc{Project} & \textsc{Commit SHA} & \textsc{Execution Time (s)} & \textsc{Execution Time (s)}\\
\midrule
maven-config-processor-plugin &  \href{https://github.com/lehphyro/maven-config-processor-plugin/tree/c92e588}{c92e588}  & \np{612}  & \np{1159}\\
lanterna  & \href{https://github.com/mabe02/lanterna/tree/5982dbfd}{5982dbf} & \np{365} & \np{1628}\\
AutoLoadCache & \href{https://github.com/qiujiayu/AutoLoadCache/tree/06f67543}{06f6754} & \np{347} & \np{5941}\\
gwt-cal & \href{https://github.com/bradrydzewski/gwt-cal/tree/e7e5250}{e7e5250} & \np{293} & \np{2075}\\
Bukkit & \href{https://github.com/Bukkit/Bukkit/tree/f210234e}{f210234} & \np{143} & \np{10452}\\
Mybatis-PageHelper & \href{https://github.com/pagehelper/Mybatis-PageHelper/tree/525394c}{525394c} & \np{128} & \np{1749}\\
\bottomrule
\end{tabular}
\end{table*}

\subsection{Threats to Validity}

\subsubsection{Internal Validity}

The threats to internal validity relate to the effectiveness of \jdbl on generic real-world Java projects, as well as the design decisions that can influence our results.
Coverage-based debloating  has some inherent limitations, \eg inadequate test cases and random behaviors. 
To mitigate these threats, we use as study subjects libraries with high coverage (see \autoref{tab:benchmark}), and execute the test suite three times  to avoid test randomness.
If the test suite does not capture all desired behaviors,  some necessary code might not be executed and  be removed.
The debloated libraries can also have non-deterministic test cases. 
For example, tests that use the current date and time to perform an action or not. 
Due to these behaviors, executing the application multiple times with the same input may lead to different coverage results. 

As explained in \autoref{sec:approach}, \jdbl relies on a complex stack of existing bytecode coverage tools.
It is possible that some of these tools may fail to instrument the classes for a particular project. 
However, since we rely on a diverse set of coverage tools, the failures of one specific tool are likely to be corrected by the others. 
\jdbl relies on the official \mv plugins for dependency management and test execution. Still, due to the variety of existing \mv configurations and plugins, \jdbl may crash at some of its phases due to conflicts with other plugins. 
To overcome this threat and to automate our experiments, we set the \texttt{maven-surefire-plugin} to its default configuration, and use the \texttt{maven-assembly-plugin} to build the fat \jar of all the study subjects.

\subsubsection{External Validity}

The threats to external validity are related to the generalizability of our findings.
Our observations in \autoref{sec:results} about bloat are made on single-module \mv projects and the Java ecosystem. Our findings are valid for software projects with these particular characteristics.
Meanwhile, coverage-based debloating in different languages could yield different conclusions than ours.
Moreover, our debloating results are influenced by the coverage of the libraries and clients used as study subjects.  
However, we took care to select open-source Java libraries available on GitHub, which cover projects from different domains (\eg, logging, database handling, encryption, IO utilities, metaprogramming, networking).\footnote{\scriptsize{See \url{https://github.com/castor-software/jdbl-experiments/blob/master/dataset/data/jdbl_dataset.json}}}
To the best of our knowledge, this is the largest set of study subjects used in software debloating experiments.

\subsubsection{Construct Validity}

The threats to construct validity are related to the relation between the coverage-based debloating approach and the experimental protocol. 
Our analysis is based on a diverse set of real-world open-source Java projects, with minimal modifications to run \jdbl (only the \textit{pom.xml} file is modified). 
We assume that all the plugins involved in the \mv build life-cycle are correct, as well as all the generated reports.
Note that, if a dependency is not resolved correctly by \mv, then its bytecode will not be instrumented. 
Thus, the quality of the debloat result  depends on the effectiveness of the \mv dependency resolution mechanism.

The applicability of coverage-based debloating   depends on the quality of the workload.
In our experiments, we rely on the projects' test suite. Consequently, our observations partly depend on the coverage of the projects.
As explained in \autoref{sec:dataset}, the coverage of the libraries in our dataset is high.
On the other hand, the coverage of the clients is lower ($ 20.24\% $ on average), which may cause some used functionalities in the debloated libraries that are not executed by the clients' tests. However, we believe this does not affect our results because we assess the semantic correctness of the client applications when using the debloated version of a library based on the client's usage intent expressed by its test suite.

\section{Related Work}\label{sec:related}

In this section, we present the works related to software debloating techniques and dynamic analysis. 

\subsection{Software Debloating}

Research interest in software debloating has grown in recent years, motivated by the reuse of large open-source libraries designed to provide several functionalities for different clients~\cite{Eder2014, Jiang2016a}.
Seminal work on debloating for Java programs was performed by Tip \etal~\cite{Tip1999, Tip2002}. 
They proposed a comprehensive set of transformations to reduce the size of Java bytecode including class hierarchy collapsing, name compression, constant pool compression, and method inlining.
Recent works investigate the benefits of debloating Java frameworks and Android applications using static analysis.
Jiang \etal~\cite{Jiang2016} presented \textsc{JRed}, a tool to reduce the attack surface by trimming redundant code from Java binaries. 
\textsc{RedDroid}~\cite{Jiang2018} and \textsc{PolyDroid}~\cite{Heath2019} propose debloating techniques for mobile devices.
They found that debloating significantly reduces the bandwidth consumption used when distributing the application, improving the performance of the system by optimizing resources.
Other works rely on debloating to improve the performance of the \mv build automation system~\cite{Ahmet2016}, removing bloated dependencies~\cite{Soto2020}, and mitigating runtime bloat~\cite{Xu2014}.
More recently, Haas \etal~\cite{Haas2020} investigate the use of static analysis to detect unnecessary code in Java applications based on code stability and code centrality measures.
Most of these works show that static analysis, although conservative by nature, is a useful technique for debloating in practice.

To improve the debloating results of static analysis, recent debloating techniques drive the removal process using information collected at runtime. 
In this context, various dynamic analysis strategies can be adopted, \eg, monitoring, debugging, or performance profiling. 
This approach allows debloating tools to collect execution paths, tailoring programs to specific functionalities by removing unused code~\cite{schultz2003automatic, Heo2018, Vazquez2019}. 
Unfortunately, most of the existing tools currently available for this purpose do not target large Java applications, focusing primarily on small C/C++ executable binaries.
Sharif \etal~\cite{Sharif2018} propose \textsc{Trimmer}, a debloating approach that relies on user-provided configurations and compiler optimization to reduce code size.
Qian \etal~\cite{Qian2019} present \textsc{RAZOR}, a tool for debloating program binaries based on test cases and control-flow heuristics. However, the authors do not provide a thorough analysis of the challenges and benefits of using code coverage to debloat software.
More recently, Bruce \etal~\cite{Bruce2020} propose \jshrink, a tool to dynamically debloat modern Java applications.
However, \jshrink is not directly automatable within a build pipeline and the effect of debloating on the library clients is not studied.
These previous works assess the impact of debloating on the size of the programs, yet, they rarely evaluate to what extent the debloating transformations preserve program behavior.

This work contributes to the state-of-the-art of software debloating. 
We propose an approach for debloating Java libraries based on the usage of code coverage to identify unused software parts.
Our tool, \jdbl, integrates the debloating procedure into the \mv build life-cycle, which facilitates its evaluation and its integration in most real-world Java projects.
We evaluate our approach on the largest set of programs ever analyzed in the debloating literature, and we provide the first quantitative investigation of the impact of debloating on the library clients. 

\subsection{Dynamic Analysis}

Dynamic analysis is the process of collecting and analyzing the data produced from executing a program. 
This long-time advocated software engineering technique is used for several tasks, such as program slicing~\cite{Agrawal1990}, program comprehension~\cite{Cornelissen2009}, or dynamic taint tracking~\cite{Bell2014}.
Through dynamic analysis, developers can obtain an accurate picture of the software system by exposing its actual behavior.
For example, trace-based compilation uses dynamically-identified frequently-executed code sequences (traces) as units for optimizing compilation~\cite{Inoue2010, Gal2009}.
Mururu \etal~\cite{Mururu2019} implemented a scheme to perform demand-driven loading of libraries based on the localization of call sites within its clients. 
This approach allows reducing the exposed code surface of vulnerable linked libraries, by predicting the near-exact set of library functions needed at a given call site during the execution.
Palepu \etal~\cite{Palepu2017} use dynamic analysis to effectively summarize the execution and behavior of modern applications that rely on large object-oriented libraries and components.
In this work, we employ dynamic analysis for bytecode reduction, as opposed to runtime memory bloat, which was the target of previous works~\cite{Mitchell2009,Nguyen2018, Mitchell2010, Guoqing2013, Xu2010, Nguyen2013, Bhattacharya2013}. 

In Java, dynamic analysis is often used to overcome the limitations of static analysis. Landman~\cite{Landman2017} performed a study on the usage of dynamic features and found that reflection was used in \np[\%]{78} of the analyzed projects. 
Recent work from Xin \etal~\cite{Xin2019} utilizes execution traces to identify and understand features in Android applications by analyzing their dynamic behavior. 
In order to leverage dynamic analysis for debloating, we need to collect a very accurate coverage report, which guides the debloating procedure.

Our work contributes to the state-of-the-art of dynamic analysis for Java programs. 
Our technique combines information obtained from four distinct code coverage tools through bytecode instrumentation~\cite{Binder2007}. 
The composition of these four types of observations allows us to build a very accurate and complete coverage report, which is necessary to identify exactly what parts of the code are used at runtime and which ones can be removed.
To collect coverage, we rely on the test suite of the libraries. 
This approach is similar to other dynamic analyses, \eg, for finding backward incompatibilities~\cite{Chen2020}.

\autoref{tab:tools_comparison} summarizes the state-of-the-art of published techniques for debloating Java applications in comparison with \jdbl. 
As observed, most existing techniques target bytecode instead of source code.  DepClean~\cite{Soto2020} is the exception, which focuses on debloating \textit{pom.xml} files based on static bytecode analysis.
\jshrink~\cite{Bruce2020} uses a combination of static and dynamic analysis to address the potential unsoundness of static analysis in the presence of new language features.
To our knowledge, \jdbl is the first fully automatic debloating technique that also debloats code in third-party dependencies, and that assesses the correctness of debloating with respect to both the successful build of the debloated library and the successful execution of the library's clients.
Furthermore, our experiments are at least one order of magnitude larger than previous works.

\begin{table*}[t]
\footnotesize
\centering
\caption{Comparison of existing Java debloating techniques, \wrt this work. 
\textsc{Target} is the type of artifact considered for debloating; 
\textsc{Analysis} refers to the type of code analysis performed for debloating; 
\textsc{Scale} provides the number of study subjects used to evaluate the technique;
\textsc{Granularity} is the code level at which debloating is performed: field (F), method (M), class (C) or dependency (D); the last columns enumerate the three ways found in the state-of-the-art to assess the validity and utility of the debloated artifact.}
\label{tab:tools_comparison}
\begin{tabular}{@{}lllp{1.4cm}cccccccc@{}}
\hline
             \multirow{2}{*}{\textsc{Tool}}               & \multirow{2}{*}{\textsc{Target}} & \multirow{2}{*}{\textsc{Analysis}} & \multirow{2}{*}{\textsc{Scale}}     & \multicolumn{4}{c}{\textsc{Granularity}}                                                                                                                                                       &  & \multicolumn{3}{c}{\textsc{Correctness evaluation criteria}}                                                   \\ \cline{5-8} \cline{10-12} 
                             &                         &                           &                                        & \multicolumn{1}{c}{\multirow{2}{*}{F}} & \multicolumn{1}{c}{\multirow{2}{*}{M}} & \multicolumn{1}{c}{\multirow{2}{*}{C}} & \multicolumn{1}{c}{\multirow{2}{*}{D}} &  & \multicolumn{1}{c}{lib}  & \multicolumn{1}{c}{lib's}  & \multicolumn{1}{c}{lib's clients} \\
                             &                         &                           &                                        & \multicolumn{1}{c}{}                        & \multicolumn{1}{c}{}                         & \multicolumn{1}{c}{}                         & \multicolumn{1}{c}{}                              &  & \multicolumn{1}{c}{compiles} & \multicolumn{1}{c}{tests pass} & \multicolumn{1}{c}{tests pass}        \\ \hline
DepClean \cite{Soto2020}     & Source                  & Static                    & $30$ libs                         & \textcolor{red}{\ding{55}}                  & \textcolor{red}{\ding{55}}                   & \textcolor{red}{\ding{55}}                   & \textcolor{teal}{\ding{51}}                       &  & \textcolor{teal}{\ding{51}}  & \textcolor{teal}{\ding{51}}    & \textcolor{red}{\ding{55}}            \\
\hline
\end{tabular}
\vspace{-0.2cm}
\end{table*}


\section{Conclusion}\label{sec:conclusion} 

In this work, we introduce coverage-based debloating for Java applications.
We have addressed one key challenge of dynamic debloating: collect accurate and complete coverage information that includes the minimum set of classes and methods that are necessary to execute the program with a given workload.
We implemented coverage-based debloating  in an open-source tool called \jdbl.
We have performed the largest empirical validation of Java debloating in the literature with \np{\nbLibCompiling} libraries and \np{\nbClient} clients that use these libraries. 
We evaluated \jdbl using an original experimental protocol that assessed the impact of debloating on the libraries' behavior, on their size, as well as on their clients.
Our results indicate that \jdbl can reduce \np[\%]{68.3} of the bytecode size and that \ChartSmall[p]{\nbLibPassTestNum}{\nbLibDebSuccessNum} debloated libraries compile and preserve their test behavior.
We also show that \jdbl outperforms \jshrink regarding size reduction and behavior preservation, when used on the same benchmark as in the \jshrink paper.
For the first time in the literature, we assess the utility of debloated libraries for their clients:  
\ShowPercentage{\nbClientDebloatNoTestFail}{\nbDynCoveringClient} of the clients  can successfully compile and run their test suite with a debloated library.

Our results provide evidence of the massive presence of unnecessary code in software applications and the usefulness of debloating techniques to handle this phenomenon.
Furthermore, we demonstrate that dynamic analysis can be used to automatically debloat libraries while preserving the functionalities that are necessary for their clients.

The next step of coverage-based debloating is to specialize applications with respect to usage profiles collected in production environments, extending the debloating to other parts of the program stack, \eg, to the Java Runtime Environment (JRE), program resources, or containerized applications. 
As for the empirical investigation of the impact of debloating, we aim to evaluate the effectiveness of coverage-based debloating in reducing the attack surface of modern applications.
These are major milestones towards full-stack debloating for software hardening.

\begin{acks}\label{sec:ak}
We would like to thank Dr. Bobby Bruce and all the authors of \jshrink for sharing the source code and providing valuable feedback during our reproduction of their experiments. This work is partially supported by the Wallenberg AI, Autonomous Systems, and Software Program (WASP) funded by Knut and Alice Wallenberg Foundation, as well as by the TrustFull and the Chains projects funded by the Swedish Foundation for Strategic Research.
\end{acks}

\bibliographystyle{ACM-Reference-Format}
\bibliography{biblio}

\end{document}